\documentclass[twocolumn]{aastex63}
\usepackage{rotating}
\usepackage{xcolor}
\usepackage{graphicx}
\usepackage{hyperref}
\usepackage{upgreek}

\usepackage{lineno}

\date{Published November 18, 2021. Accepted August 12, 2021. Received August 11, 2021; in original form December 2, 2020}

\newcommand{\chimepsr}{CHIME/Pulsar}
\newcommand{\chimefrb}{CHIME/FRB}
\newcommand{\presto}{PRESTO}
\newcommand{\psrchive}{PSRCHIVE}
\newcommand{\dspsr}{DSPSR}
\newcommand{\tempotwo}{TEMPO2}
\newcommand{\tempo}{TEMPO}
\newcommand{\dmunits}{$\mathrm{pc\,cm^{-3}}$}
\newcommand{\us}{$\upmu$s}

\newcommand{\ubc}{Department of Physics \& Astronomy, University of British Columbia, 6224 Agricultural Road, Vancouver, BC V6T 1Z1, Canada}
\newcommand{\mcgilldep}{Department of Physics, McGill University, 3600 rue University, Montr\'eal, QC H3A 2T8, Canada}
\newcommand{\mcgillsi}{McGill Space Institute, McGill University, 3550 rue University, Montr\'eal, QC H3A 2A7, Canada}
\newcommand{\mitkavli}{MIT Kavli Institute for Astrophysics and Space Research, Massachusetts Institute of Technology, 77 Massachusetts Ave, Cambridge, MA 02139, USA}
\newcommand{\mitdep}{Department of Physics, Massachusetts Institute of Technology, 77 Massachusetts Ave, Cambridge, MA 02139, USA}

\newcommand{\dunlap}{Dunlap Institute for Astronomy \& Astrophysics, University of Toronto, 50 St. George Street, Toronto, ON M5S 3H4, Canada}

\begin{document}

\title{First discovery of new pulsars and RRATs with CHIME/FRB}
\correspondingauthor{Deborah C. Good}
\email{deborah.good@nanograv.org}

\author[0000-0003-1884-348X]{D.~C.~Good}
\affiliation{\ubc}

\author[0000-0001-5908-3152]{B.~C.~Andersen}
\affiliation{\mcgilldep}
\affiliation{\mcgillsi}

\author[0000-0002-3426-7606]{P.~Chawla}
\affiliation{\mcgilldep}
\affiliation{\mcgillsi}

\author[0000-0002-1529-5169]{K.~Crowter}
\affiliation{\ubc}

\author[0000-0003-4098-5222]{F.~Q.~Dong}
\affiliation{\ubc}

\author[0000-0001-8384-5049]{E.~Fonseca}
\affiliation{\mcgilldep}
\affiliation{\mcgillsi}
 \affiliation{Department of Physics and Astronomy, West Virginia University, P.O. Box 6315, Morgantown, WV 26506, USA}
 \affiliation{Center for Gravitational Waves and Cosmology, West Virginia University, Chestnut Ridge Research Building, Morgantown, WV 26505, USA}

\author[0000-0001-8845-1225]{B.~W.~Meyers}
\affiliation{\ubc}

\author[0000-0002-3616-5160]{C.~Ng}
\affiliation{\dunlap}

\author[0000-0002-4795-697X]{Z.~Pleunis}
\affiliation{\mcgilldep}
\affiliation{\mcgillsi}

\author[0000-0001-5799-9714]{S.~M.~Ransom}
\affiliation{National Radio Astronomy Observatory, 520 Edgemont Rd., Charlottesville, VA 22903, USA}

\author[0000-0001-9784-8670]{I.~H.~Stairs}
\affiliation{\ubc}

\author[0000-0001-7509-0117]{C.~M.~Tan}
\affiliation{\mcgilldep}
\affiliation{\mcgillsi}

\author[0000-0002-3615-3514]{M.~Bhardwaj}
\affiliation{\mcgilldep}
\affiliation{\mcgillsi}

\author[0000-0001-8537-9299]{P.~J.~Boyle}
\affiliation{\mcgilldep}
\affiliation{\mcgillsi}
\affiliation{Department of Physics and Astronomy, West Virginia University, P.O. Box 6315, Morgantown, WV 26506, USA}
\affiliation{Center for Gravitational Waves and Cosmology, West Virginia University, Chestnut Ridge Research Building, Morgantown, WV 26505, USA}

\author[0000-0001-7166-6422]{M.~Dobbs}
\affiliation{\mcgilldep}
\affiliation{\mcgillsi}
\affiliation{University of Oxford, Sub-Department of Astrophysics, Denys Wilkinson Building, Keble Road, Oxford, OX1 3RH, United Kingdom}

\author[0000-0002-3382-9558]{B.~M.~Gaensler}
\affiliation{\dunlap}
\affiliation{David A. Dunlap Department of Astronomy \& Astrophysics, University of Toronto, 50 St. George Street, Toronto, ON M5S 3H4, Canada}

\author[0000-0001-9345-0307]{V.~M.~Kaspi}
\affiliation{\mcgilldep}
\affiliation{\mcgillsi}

\author[0000-0002-4279-6946]{K.~W.~Masui}
\affiliation{\mitkavli}
\affiliation{\mitdep}

\author[0000-0002-9225-9428]{A.~Naidu}
\affiliation{\mcgilldep}
\affiliation{\mcgillsi}

\author{M.~Rafiei-Ravandi}
\affiliation{Perimeter Institute for Theoretical Physics, 31 Caroline Street N, Waterloo ON N2L 2Y5 Canada}

\author[0000-0002-7374-7119]{P.~Scholz}
\affiliation{\dunlap}

\author{K.~M.~Smith}
\affiliation{Perimeter Institute for Theoretical Physics, 31 Caroline Street N, Waterloo ON N2L 2Y5 Canada}

\author[0000-0003-2548-2926]{S.~P.~Tendulkar}
\affiliation{Department of Astronomy and Astrophysics, Tata Institute of Fundamental Research, Mumbai, 400005, India}
\affiliation{National Centre for Radio Astrophysics, Post Bag 3, Ganeshkhind, Pune, 411007, India}

\begin{abstract}
    We report the discovery of seven new Galactic pulsars with the Canadian Hydrogen Intensity Mapping Experiment's Fast Radio Burst backend (\chimefrb{}). These sources were first identified via single pulses in \chimefrb{}, then followed up with \chimepsr{}. Four sources appear to be rotating radio transients (RRATs), pulsar-like sources with occasional single pulse emission with an underlying periodicity. Of those four sources, three have detected periods ranging from 220 ms to 2.726 s. Three sources have more persistent but still intermittent emission and are likely intermittent or nulling pulsars. We have determined phase-coherent timing solutions for the latter \replaced{three}{two}. These seven sources are the first discovery of previously unknown Galactic sources with \chimefrb{} and highlight the potential of fast radio burst detection instruments to search for intermittent Galactic radio sources. 
\end{abstract}

\keywords{Pulsars (1306), Radio pulsars (1353)}

\section{Introduction}
In the five decades since their initial discovery, radio pulsars (rapidly rotating, highly magnetized neutron stars) have proven to be excellent tools to study a wide variety of astrophysics and physics, including tests of general relativity, and they are also fascinating in their own right. To date, astronomers have discovered more than 2,800 pulsars in the Milky Way and Magellanic Clouds \citep{mht+05}\footnote{See \url{http://www.atnf.csiro.au/people/pulsar/psrcat.}}. Most pulsars emit consistently at radio wavelengths, but a subset emit only intermittently or experience nulling. Nulling pulsars experience sharp, sudden drops in pulse energy for a few periods, as first reported by \cite{b70}.
Intermittent pulsars are an intermediate category, displaying intermittent emission but behaving as a conventional pulsar during active periods \citep{l09}. In 2006, careful re-examination of Parkes Multibeam Pulsar Survey data \citep[PKMBS; ][]{mlc+01} introduced Rotating Radio Transients (RRATs) to the field---sources with occasional pulsar-like emission and underlyings periodicity \citep{mll+06, mlk+09}. RRATs remain somewhat loosely defined as pulsar-like sources detected via single pulses instead of a periodicity search, and may exist as part of a continuum including nulling and intermittent pulsars \citep{b+13}. Intermittent pulsars are distinguished observationally from RRATs by whether detections can be improved by folding at a known period.

Pulsar search strategies are well established, usually falling into two categories: blind searches and targeted searches. In blind searches, a specified region of sky is searched for pulsar emission. In targeted searches, a region of interest such as a known supernova remnant, globular cluster, or other likely pulsar environment is selected and thoroughly searched, as in, e.g., \cite{rhs+05}. Modern pulsar searches such as the Parkes High Time Resolution Universe (HTRU), the recently terminated Arecibo L-Band Feed Array Pulsar Survey (PALFA), and the Green Bank North Celestial Cap (GBNCC) survey are highly effective at discovering both ``slow'' and millisecond pulsars \citep{kjv+10,cfl+06,slr+14}.

These surveys, however, are limited in the amount of time they can spend focusing on each pointing, as most such surveys make use of steerable telescopes with limited time allocations. This poses two relevant challenges. First, surveys are susceptible to missing intermittent pulsars or pulsars with a significant nulling fraction. Second, pulsar searching programs are sometimes unable to conduct sufficient follow-up observations to determine complete timing solutions.

The Canadian Hydrogen Intensity Mapping Experiment (CHIME) has several backends enabling a wide-range of science. Among those, the CHIME Fast Radio Burst (\chimefrb) and CHIME Pulsar Timing (\chimepsr) systems are uniquely situated to detect and time intermittent or otherwise unusual sources which produce bright single pulses.  The sister systems were primarily designed to separately discover FRBs and to time known pulsars, but when combined, the search engine can identify bright single pulses from Galactic sources and the pulsar instrument can collect search-mode observations to find initial timing solutions then daily fold-mode observations to improve timing solutions. 
\deleted{Though detections of FRBs in pulsar searches are common (\citealp[e.g.][]{lbm+07,cpk+16,pab+18}), this work is among the  first discovery of pulsars detected via single pulses in an FRB search.}

In this work, we discuss the discovery of seven new sources, including timing solutions where possible. In Section~\ref{sec:chime_overview}, we provide a brief overview of CHIME and the \chimefrb{} and \chimepsr{} systems. In Section~\ref{sec:detection}, we describe our method for detecting and characterizing new Galactic sources from single pulses, including  discussion of the possibility of future detections. In Sections~\ref{section:source_discussion} and \ref{sec:conclusion}, we discuss the individual sources and their interpretation, with a focus on the future potential of this method. 

\section{CHIME Overview}\label{sec:chime_overview}
CHIME is a transit radio telescope, located at the Dominion Radio Astrophysical Observatory in British Columbia, Canada. It consists of four cylinders, each 100\,m by 20\,m, with only 80\,m of their length illuminated and with 256 dual-polarization antennae on each focal line. As a transit telescope, CHIME views the sky from declination $+90 \degr$ to roughly $-20 \degr$ between 400--800 MHz every day for roughly fifteen minutes.  

CHIME is a powerful instrument for a variety of science goals, including mapping redshifted neutral hydrogen emission, detecting FRBs, and timing pulsars. Separate backends allow these observations to be entirely commensal.

\subsection{The \chimefrb{} System \label{frboverview}}
The \chimefrb{} backend \citep{chimefrb} is a dispersed millisecond radio transient discovery instrument, which forms 1,024 static beams, tiling the approximately 120-degree north-south and 2-degree east-west CHIME primary beam. The beams are formed using a combination of brute force and Fast Fourier Transform beamforming \citep[see e.g.,][]{nvp+17, msn+19}.  Each beam has a frequency-dependent $20-40'$ full width at half maximum (FWHM) diameter \citep{chimefrb}.  
CHIME/FRB data \citep{chimefrb} require immediate dedispersion and searching, as the volume of data passing through the system is too large to be saved. 
The backend uses a tree dedispersion model, a concept described generally in \replaced{\cite{lk04}}{ \cite{taylor74}} and briefly in a \chimefrb{} context in \cite{chimefrb}, to dedisperse \replaced{data. and events are classified as ``Galactic'', ``Ambiguous,'' or ``Extragalactic'' based on their dispersion measure (DM). Triggers on the same event from different beams are grouped, and radio frequency interference (RFI) is excised throughout the process. For events which pass through the CHIME/FRB searching backend and appear to be astrophysical and are labeled ``Extragalactic'', buffered intensity data is written to disk, saving data from specific triggers.}

{ 
Events are classified based on whether their DM indicates they are extragalactic. We determine the maximum Galactic DM in the direction of the candidates using both the \cite{cl02} and \cite{ymw17} Galactic DM maps and calculate the difference between the two estimates. We then add in quadrature the uncertainties in measured DM and the difference between the two maximum Galactic DM estimates. This is our total uncertainty $\sigma$. Sources where the measured DM exceeds both the maximum Galactic DM models by at least $5\sigma$ are classified as ``Extragalactic.'' If the measured DM is $2 \sigma$ to $5 \sigma$ in excess of the maximum Galactic DM, the source is classified as Ambiguous. If the measured DM is less than $2 \sigma$ in excess of the maximum Galactic DM, the source is classified as Galactic \citep{chimefrb}.}


Though the \chimefrb{} system  is fully configurable, it is optimized for finding extragalactic FRBs, so the \chimefrb{} team chooses to apply a DM cut, saving intensity data to disk only for events labeled as ``Extragalactic'' and ``Ambiguous'' DM. This means that, although we record the existence of triggers with Galactic DM in the \chimefrb{} database, we do not generally record intensity data. However, metadata, including timestamps and signal-to-noise ratios (S/N), are saved for all CHIME/FRB events regardless of DM. The value of \chimefrb{} metadata are explored in more depth in Section \ref{headeranalysis}. Additionally, we can request future saved intensity data from specific sources once they have been identified by \chimefrb{} metadata. 

Once a new Galactic source is detected with \chimefrb, it can be studied in more detail by observing it with \chimepsr{}.

\subsection{The \chimepsr{} System \label{psroverview}}

The \chimepsr{} backend \citep{chimepsr} is a dedicated pulsar timing instrument that uses 10 independent tied-array beams formed in real time by the CHIME correlator. In contrast to the static \chimefrb{} beams, \chimepsr{} beams digitally track sources and are therefore more consistently sensitive than \chimefrb{} beams. The backend can produce coherently dedispersed ``fold-mode" and ``search-mode" data formats, with the 400-800\,MHz band sampled by 1,024 frequency channels. When following up \chimefrb{} single pulse candidates, we use the search-mode (filterbank) format, recording a coherently dedispersed total intensity time series with 327.68-\us{} resolution for each of the 1,024 frequency channels. Once a basic timing solution is available, it is possible to switch future observations to generating folded data products which can then be used to further refine the candidate ephemeris with near-daily observations. This final step is only performed for pulsars that show detectable persistent emission. 

\section{Detection, timing, and analysis methods}\label{sec:detection}
\subsection{Initial detection}

As \chimefrb{} is a passive monitor, it will detect any emission from a source as it transits. Each of the seven sources presented in this work was a serendipitous discovery, not the result of a targeted search. \chimefrb{} scientific personnel acting as system monitors manually identified these sources as not corresponding to any known Galactic sources and flagged them for follow-up. \added{This was not a systematic process and therefore it is not possible to estimate its completeness}. However, we do realise the value for a thorough search through the \chimefrb{} database; in a subsequent work we will present a detailed method and the results for a \added{systematic search of the \chimefrb{} database}. The pulsars discovered in this alternate method have already been announced on the \chimefrb{} public galactic webpage.\added{\footnote{New Galactic pulsars discovered by \chimefrb{} can be viewed at \url{https://www.chime-frb.ca/galactic}.}} 

These sources have been primarily identified using metadata detections, without intensity data. However, after initial detection, we manually configure the \chimefrb{} system to produce a small set of saved intensity data for each source to confirm that they are astrophysical and not RFI \added{by examining a frequency vs. time ``waterfall'' plot of the \chimefrb{} intensity data. This intensity data are used only to confirm that sources are astrophysical and do not contribute to timing solutions.}

\subsection{\chimefrb{} metadata analysis \label{headeranalysis}}
As metadata are available for all events detected by \chimefrb{}, they provide a daily register of single pulses from Galactic sources, and can help characterize new Galactic sources discovered with \chimefrb{}.

We have queried the CHIME/FRB database for all events with S/N $>$ 8 associated with the sources presented here between 2018-08-28 (MJD 58358) up to and including 2020-05-01 (MJD 58970) to create per-source sets of data. \added{The \chimefrb{} system was operating nominally for 596 days during this period.}

\begin{figure*}
    \centering
    \includegraphics[width=\textwidth]{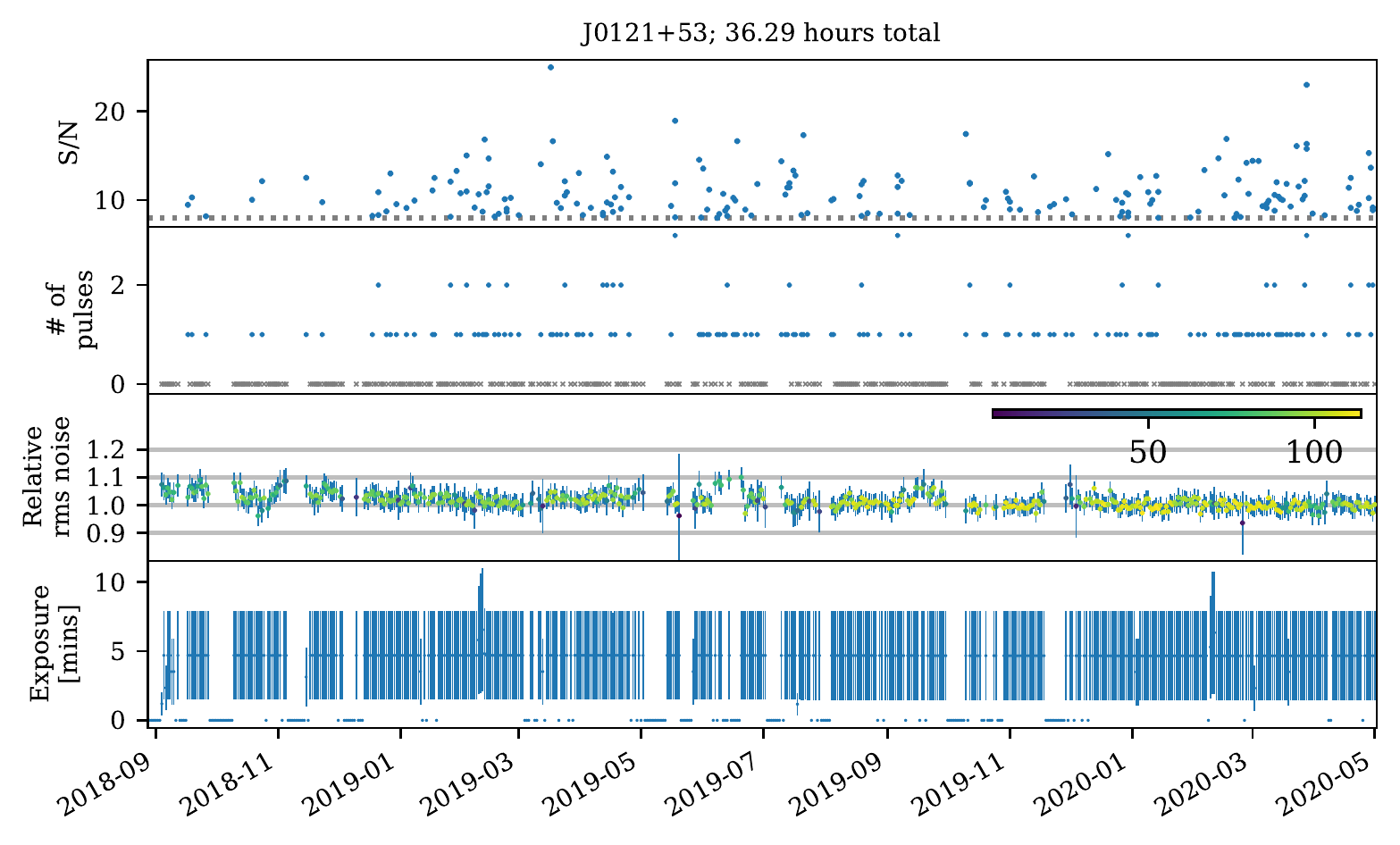}
    \caption{CHIME/FRB detections of new RRAT-like source PSR J0121+53, shown as an example, along with sensitivity to this location and exposure time at this location.  The top panel shows the daily median detection S/N and standard deviation of the in-beam events with S/N $>$ 8. The second panel shows the number of single pulses that were detected. If no pulse was detected on a day with non-zero exposure the marker is a gray cross. The third panel shows the daily sensitivity of the experiment as characterized by the relative rms noise of known pulsar detections. The color bar in the inset shows how many pulsars were used to to obtain the average rms noise. The bottom panel shows the exposure to the source's localization uncertainty region. \added{On some days the exposure time is larger than its typical value because of the occurrence of two transits in the same Solar day.}}
    \label{fig:cf_detections}
\end{figure*}

To analyze source activity with \chimefrb{} metadata, we generally keep only the events that were detected when a source was within the FWHM at 600 MHz of any of the CHIME/FRB formed beams, where our understanding of the exposure and system sensitivity is best. \added{\chimefrb{} beam widths are frequency dependent, and the FWHM of the beams do not overlap at all frequencies. (See  \cite{chimefrb} for a more detailed discussion of beam placement).  This means that some sources may fall between beams at higher frequencies.} \replaced{As J0209+5759 falls outside of the formed beams at 600 MHz, we consider the FWHM at 400 MHz, only for this source.}{This is the case for J0209+5759 at 600 MHz. Therefore, we consider the FWHM at 400 MHz, for this source alone}. We show the CHIME/FRB detections of J0121+53 as an example in Figure~\ref{fig:cf_detections}, demonstrating the breadth of information that can be gleaned from \chimefrb{} metadata. \added{\footnote{Versions of this figure are available for the other six sources at \url{https://www.chime-frb.ca/galactic}.}} This figure shows the S/N for each \chimefrb{} detection, the number of pulses seen per day by \chimefrb{}, the daily relative system sensitivity, and the daily exposure time for the location of each source. \replaced{Uncertainties on the daily exposure times represent the standard deviations among exposure times in the uncertainty regions of each source.}{We determine uncertainties for daily exposure times by first determining the uncertainty region of each source's position, then measuring the daily exposure time in the extent of that region. We calculate the standard deviation of these values to determine the 68\% confidence interval for the exposure time.}

\begin{figure}
    \centering
    \includegraphics[width=\columnwidth]{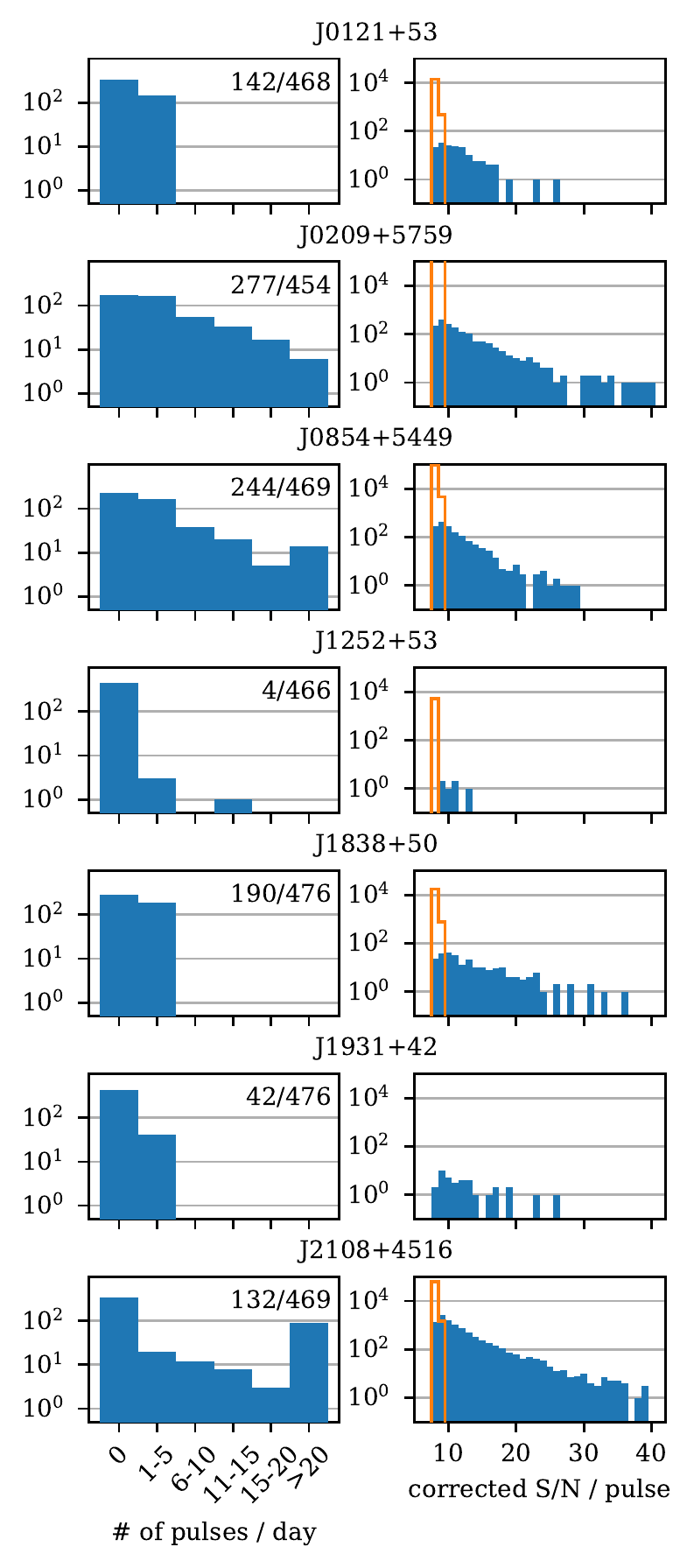}
    \caption{Summary of CHIME/FRB detections of new pulsars and RRATs within one of the synthesized beams. Left: number of pulses detected in each \replaced{transit}{Solar day}, with the number of days with at least one detection as well as the total number of days with exposure in the top right corner. Right: the S/Ns of all single pulses (blue) and upper limits from non-detections (estimated from the spin period and transit duration; orange) divided by the daily root mean square (rms) system sensitivity as determined from detections of known pulsars.}
    \label{fig:cf_detections_histograms}
\end{figure}

For all sources presented here we have compiled the number of detections per day and the S/N per detection in Figure~\ref{fig:cf_detections_histograms}. In this analysis, we have included only days with non-zero exposure when the system was functioning nominally. Using the spin period of the sources (see Tables \ref{tab:period_determination} and \ref{tab:parameter_table}), we have estimated the number of missed pulses in each transit and assign missed pulses the S/N threshold of 8, used as the upper limit for the non-detection. We correct the detection S/Ns and upper limits for day-to-day sensitivity variations by dividing them by the per-day root mean square (rms) sensitivity variation, as derived from monitoring known pulsars.

\replaced{As \chimefrb{} is not intentionally scheduled, }{Nominally, \chimefrb{} observations occur for all observable regions of the sky every day, which makes} \chimefrb{} metadata \deleted{are} an ideal source of burst rate estimates. We use existing exposure time calculations and complete sets of \chimefrb{} metadata detections to determine a basic burst rate for each source and treat the uncertainty in this rate calculation as Poissonian. For this calculation we count bursts and calculate the exposure at the best-known sky positions of each source, without marginalizing over its uncertainty region (see Table \ref{tab:parameter_table}).

Including all source events from the database, we convert the metadata to time-of-arrival (TOA) files formatted for common pulsar timing software suites. TOAs generated from CHIME/FRB metadata have relatively large uniform DM uncertainties, $1.62 \times 2^i$\,\dmunits{}, and uniform TOA uncertainties, $\sim 31.5 \times 2^i$\,ms, where $i \in \{0, 1, 2, 3, 4\}$ is the tree index. Pulses are optimally detected in the tree index that most closely matches their pulse width. These uncertainties are determined by the coarse-graining step in the tree dedispersion algorithm that takes the maximum S/N over each $64 \times 64$ block of fine-grained search trials \citep{chimefrb}.\footnote{Coarse-graining factors are configurable and are lower as of late 2020.} 

We estimate pulsar and RRAT spin periods from the metadata TOAs by brute-force searching for the \replaced{greatest common denominator}{spin period which yields the greatest integer number of periods in}
 the time intervals between bursts, using multiple days of data. To overcome a bias towards detecting multiples of the coarse-grained sampling time as the spin period, we include Monte Carlo resampling, wherein every iteration all TOAs are placed at a random position within their uniform uncertainty region before the spin period estimation. The goodness-of-fit metric of all instances is averaged together to find a global best spin period estimate. This method's efficacy was confirmed by blindly redetecting the spin periods of known pulsars from metadata TOAs.

An initial coarse timing solution can be derived from the CHIME/FRB metadata TOAs using standard pulsar timing software such as \tempo{} or \tempotwo{}.  Though this solution has large residuals when compared to conventional pulsar timing solutions, it generates reliable starting solutions, \replaced{to refine \chimepsr{} initial single pulse solutions and enable fold-mode observations with \chimepsr{}}{phase-connected within observations and over longer-periods of time. These solutions enable us to collect fold-mode observations with \chimepsr{} for sources that demonstrate persistent emission and provide a valuable starting point for \chimepsr{} timing solutions for both persistent and RRAT-like sources.} 

\subsection{\chimepsr{} data analysis \label{section:chime_psr}}
After candidate sources are detected with \chimefrb{}, they are added to the \chimepsr{} automated observing scheduler described in \cite{chimepsr} to obtain more sensitive observations. \chimepsr{} initially observes sources in search-mode, as discussed in Section \ref{section:chime_psr}. The exact duration of these observations varies, but ranges between 1,400 and 1,800 seconds. We examine this search mode data using PRESTO \citep{presto} as described in Sections \ref{section:sp_detection} and \ref{section:period_search}. Once we have discovered a reliable spin period as described in Section \ref{section:period_search}, we can collect fold-mode data with \chimepsr{}.

We use a combination of search-mode and fold-mode data from \chimepsr{} to constrain properties of these new sources, including flux density and width as discussed in Sections \ref{section:flux_density} and \ref{section:width}. Where possible, we also construct timing solutions as discussed in \ref{section:timing_procedure}.

\subsubsection{RFI Excision \label{section:rfi_excision}}
We begin by removing a list of common known corrupted frequency channels in the CHIME frequency band with the PSRCHIVE \citep{psrchive04, psrchive12} \texttt{paz} routine; this excises $\sim$15\% of the band.

Before searching for single pulses, we use the PRESTO \texttt{rfifind} tool to determine individualized RFI masks (in addition to the known bad channels) for search-mode data. We also construct a common list of periodic RFI signals in \chimepsr{} data, and apply that list when appropriate. 

To remove RFI from folded data, we again employ the PSRCHIVE \texttt{paz} routine, removing the known bad channels and allowing automated excision by using the built-in median difference filter algorithm. In some cases further RFI excision was conducted using the interactive \texttt{pazi} tool \added{in PSRCHIVE}.

\subsubsection{Single pulse detections \label{section:sp_detection}}
We analyze search-mode data from \chimepsr{} using PRESTO tools, searching for both single pulse detections and periodicity. Apparent single pulse detections are then examined in more detail using PSRCHIVE tools, the PRESTO waterfaller tool, or both to confirm detection. Once a source is confirmed, initial single pulse detections are also added to the public CHIME/FRB pulsar detection webpage.

\subsubsection{Period Search \label{section:period_search}}
Where possible, we use PRESTO's \texttt{accelsearch} function to search for the candidate's period. Where \texttt{accelsearch} failed, we have also used fast folding algorithm (FFA) methods from \cite{mbs+20} to search for periodicity. However, several of these sources are likely RRATs, making it challenging to robustly detect periodicity using standard pulsar tools. 

In these cases, we also make use of brute force methods, in particular PRESTO utilities \texttt{single\_pulse\_search} and \texttt{rrat\_period}, to determine periods. For comparison purposes, we also calculate spin periods using \chimefrb{} metadata as described in Section \ref{headeranalysis} where possible. The input to \texttt{rrat\_period} is a list of times, generally the times at which the pulsar was detected within the span of one observation. It then computes intervals between those times and calculates the period which makes the number of periods in each interval as close to an integer as possible. However, our sources are highly intermittent and often \texttt{single\_pulse\_search} only detected two or three pulses within an observation. 

To determine the best possible period estimates for these highly intermittent pulsars, we made a slight modification to \texttt{rrat\_period}, as \texttt{rrat\_period\_multiday}, which allowed us to incorporate detections on multiple days. This utility calculates intervals between times which occur within each observation, but ignores the gap between observations, as is also the case in our \chimefrb{} metadata period searches. The procedure for determining a period from this point on is unchanged from \texttt{rrat\_period}. Previous investigations by \cite{cbm+17} suggest that such brute force methods are 99\% accurate with \replaced{eight intervals}{nine pulses}, but substantially less accurate with fewer pulses. Most of the sources presented here \replaced{have fewer than eight intervals}{do not meet that standard}, so while the method is robust, individual period values may not be. With more intervals available, a more accurate period is found, assuming that the source's period is approximately constant over the course of the observations being combined. \texttt{rrat\_period\_multiday} has been integrated into PRESTO.

As all observations are conducted within a narrow window in hour angle, we cannot be certain that we have not captured a sidereal-day sampling alias with CHIME/Pulsar, i.e., the detected spin frequency  $f_{\rm det} = f_0 + n/t_{\rm sidereal}$\,Hz, where $f_0$ is the true frequency, $n$ is a small integer and $t_{\rm sidereal}$ is the duration of a sidereal day in seconds. This degeneracy could be broken by additional observations with instruments other than CHIME.

\subsubsection{\chimepsr{} flux density estimation \label{section:flux_density}}
The illuminated collecting area of CHIME is approximately $\rm 80\,m\times80\,m = 6400\,m^2$.
With a typical efficiency of $\sim 50\%$ this corresponds to an effective gain $G\approx 1.16\,\mathrm{K\,Jy^{-1}}$.
Theoretically, this is also the gain for a tracking (tied-array) beam, ignoring imperfections in the beamforming and phase-calibration process.
The CHIME primary beam attenuates signals at large zenith angles and varies with frequency, thus a beam correction is also applied to derive an effective gain.
We compute the gain for the mid-point of the target observation (i.e., when the pulsar candidate is within the most sensitive region of the primary beam) and only at 600\,MHz.
The nominal contribution to the system temperature from the receiver electronics and telescope structure is $T_{\rm rec}\approx 30$\,K.
For each pulsar, the sky temperature contribution ($T_{\rm sky}$) is calculated at the nominal position from the 408\,MHz Haslam map \citep{hss+82,rdb+15}, and scaled to 600\,MHz assuming that the sky contribution scales as $\nu^{-2.5}$ \citep[e.g.,][]{oat+08}.

For each observation the band-averaged peak flux density is given by the standard radiometer equation
\begin{equation}
    S_{\rm peak} = \frac{T_{\rm rec} + T_{\rm sky}}{G\sqrt{n_{\rm p}\Delta\nu \Delta t}} \times {\rm S/N},
    \label{eq:snr_to_flux}
\end{equation}
where $n_{\rm p}=2$ is the number of polarization streams summed, $\Delta\nu$ is the effective bandwidth in Hz, and $\Delta t$ is the \replaced{integration}{sampling} time in seconds.

To estimate the mean flux density (where appropriate) we assume $S_{\rm mean} = \delta S_{\rm peak}$, where the duty cycle \replaced{$\delta = A/(S_{\rm peak})$}{$\delta = A/(PS_{\rm peak})$}, $A$ is the area under the pulse profile, \added{and $P$ is the pulse period}.
The effective bandwidth is typically $\sim 330$\,MHz after excising frequency channels that are persistently corrupted with RFI.
Preliminary instrumental evidence suggests that the system temperature may be an underestimate, perhaps by a factor of two to three. Investigations into this apparent discrepancy are ongoing. Therefore, the flux density values presented here should be considered lower limits only.

\subsubsection{Pulse Widths \label{section:width}}
To measure the pulse widths, we fit a single Gaussian profile to each single pulse (or integrated profile where available) using a non-linear least squares approach.\footnote{Specifically, we used the {\tt scipy.optimize.curve\_fit} function.}

The pulse widths are taken to be the full-width-at-half-maximum calculated of the best-fitting Gaussian, i.e., $W_{50} = 2\sqrt{2\ln 2}\sigma$, where $\sigma^2$ is the best-fit variance.
It is important to note that the number of profile phase bins can affect this measurement, and that all profiles were downsampled to a consistent 256 phase bins.

\subsubsection{Timing Solutions \label{section:timing_procedure}}
When we are able to find a spin period, our next step is to attempt to build the best possible timing solution. We begin by attempting to fold search-mode data at the source's nominal period, using the PRESTO \texttt{prepfold} routine and Digital Signal Processing Software for Pulsar Astronomy \citep[\dspsr{};][]{dspsr}. 

Depending on the sources' response to folding, our timing procedure bifurcates. \replaced{About half}{Three} can be folded using their nominal periods. For these sources, discussed in Section \ref{section:persistent_sources}, we follow a standard pulsar timing procedure, folding all search-mode data using \dspsr{} and excising RFI and generating TOAs with \psrchive{} routines \texttt{paz} and \texttt{pat} respectively. We use a standard profile created with \psrchive{} routine \texttt{paas} with a representative observation. We then determine timing solutions using \tempo{} \citep{tempo} and \tempotwo{} \citep{tempo2a}. We also shift to collecting fold-mode data for \replaced{these}{persistent} sources. 

\replaced{Other}{The other four} sources are sufficiently RRAT-like that folding the data does not result in an improved detection. For these sources, discussed in Section \ref{section:SHI}, we use \dspsr{} to create single pulse archives for each detection, generate a TOA for each single pulse, and use \tempo{} and \tempo2{} to refine our initial single pulse period solution. Where available, we also make use of \chimefrb{} metadata solutions as starting points for a timing model.

Table \ref{tab:period_determination} \replaced{outlines}{lists} the period determination methods used for each source; methods used for a given source are designated by ``Yes'' and methods not used are designated by ``No.'' Due to the minimal data present for \replaced{these}{the four RRAT-like} sources, we do not \replaced{attempt to fit a timing solution beyond refining period estimates}{fit timing parameters besides the period.}\added{ For the period, we fit with \tempo{}, using single pulsar archives where folded archives are not available.} Uncertainties in these periods are determined from \added{this} \tempo{} fitting. However, these uncertainties should be regarded as substantial underestimates; these solutions are based on a small number of TOAs which are created from single pulse archives instead of folded archives.  Where possible, we compare \chimepsr{} periods with \chimefrb{}  metadata periods. Though both use data collected by CHIME, the \chimefrb{} and \chimepsr{} provide near independent determinations of spin period.

\begin{sidewaystable}[htbp]
\centering
\caption{Period determination methods for each of the seven sources. 
\label{tab:period_determination}}
\begin{tabular}{lcccccc}
\hline\hline
Pulsar &\chimefrb{} \tablenotemark{a}& Single Pulse & Folded & $N_{toas}$\tablenotemark{b} &  Period\tablenotemark{c} & DM \\
Name  & metadata & TOAs & TOAs & & Determination & Determination \\
\hline
PSR J0121+53    & Yes  & Yes  & No  & 11 & \tempo{} & $S/N$ maximization \\
PSR J0209+5759  & Yes & No & Yes & 136 & \tempo & \texttt{prepfold} search \\
PSR J0854+5449  & Yes & No & Yes & 1,909 \tablenotemark{e} & \tempo & Subbanded timing \\
PSR J1252+53    & No & Yes & No & 10 & \tempo & $S/N$ maximization \\
PSR J1838+50    & Yes & Yes & No & 7 & \texttt{rrat\_period\_multiday} & $S/N$ maximization \\
PSR J1931+42    & No & No & No & -- & -- & $S/N$ maximization\\
PSR J2108+4516  & Yes & No & Yes & 124 & \tempo{} & PulsePortraiture \\ 
\hline
\end{tabular}
\tablenotetext{a}{See Section \ref{headeranalysis} for a discussion of \chimefrb{} metadata timing.}
\tablenotetext{b}{$N_{\rm toas}$ is the number of TOAs used in the single single pulse TOA TEMPO analysis for RRAT-like sources or the  folded TOA TEMPO analysis, depending on which is present for the source.}
\tablenotetext{c}{See Section \ref{section:period_search} for a description of \texttt{rrat\_period\_multiday}. See Section \ref{section:timing_procedure} for a discussion of \tempo-based timing with single pulse and folded archives.}
\tablenotetext{d}{See Section \ref{section:source_discussion} for an expanded discussion of DM determination.}
\tablenotetext{e}{The TOA count for J0854+5449 is based on subbanded TOAs, instead of single TOAs per observation.}
\end{sidewaystable}

\section{New sources detected with \chimefrb \label{section:source_discussion}}
For all sources, we report the source's position, DM, rotation period, and burst rate in Table \ref{tab:parameter_table} and flux density in Table \ref{flux_table}. Uncertainties are 68\% confidence intervals. 

\replaced{
DM, period, flux density, and the positions of PSR J0209+5759 and PSR J0854+5449  are derived from \chimepsr{} data. For PSR J0854+5449, DM uncertainty is determined from the subbanded timing solution. For PSR J0209+5759, the DM uncertainty is determined based on \texttt{prepfold} DM search step size. For PSR J2108+4516, DM uncertainty is found with PulsePortraiture DM fitting \citep{p19}. For highly intermittent sources where folding is not possible, DM uncertainty is determined by finding the signal-to-noise maximizing DM during each detection and averaging. Period uncertainties are determined as outlined in Section \ref{section:period_search}, and flux densities are determined as outlined in Section \ref{section:flux_density}. \chimepsr{} positions and their uncertainties are based on timing solutions.

Burst rates and positions for all other sources are determined from \chimefrb{} data, as discussed in Section \ref{headeranalysis}. 

}{Periods and flux densities for all sources are derived from \chimepsr{} data, as are the positions of PSR J0209+5759 and PSR J0854+5449. Period uncertainties are determined as outlined in Section \ref{section:period_search}, and flux densities are determined as outlined in Section \ref{section:flux_density}. \chimepsr{} positions and their uncertainties are based on timing solutions. Burst rates and all other positions are determined from \chimefrb{} data, as discussed in \ref{section:timing_procedure}. 

DM uncertainties are determined uniquely for each source. For PSR J0854+5449, DM uncertainty is determined from the subbanded timing solution. For PSR J0209+5759, the DM uncertainty is determined based on \texttt{prepfold} DM search step size. For PSR J2108+4516, DM uncertainty is found with PulsePortraiture DM fitting \citep{p19}. For highly intermittent sources where folding is not possible, DM uncertainty is determined by finding the signal-to-noise maximizing DM during each detection and averaging. }

\begin{sidewaystable}[hbtp]
\centering
\caption{Basic parameters for all new sources.
\label{tab:parameter_table}}
\begin{tabular}{lccccccc}
\hline\hline
Pulsar name & RA (hms) & Dec (dms) & DM (\dmunits) & $D_{\rm NE2001}$ (kpc) & Pulse Period \tablenotemark{a} (s) & Pulse width\tablenotemark{b} (ms) & Burst rate\tablenotemark{c} (hr$^{-1}$)  \\
\hline
PSR J0121+53   & $01^{\rm h} 21^{\rm m} \pm 11^{\rm m}$  & $+53\degr{} 29' \pm 16'$  & $91.38(3)$ & $3.3_{-0.7}^{+1.0}$    & $2.7247846(4)$ & $20(5)$ & $2.4(2)$ \\
PSR J0209+5759 & $ 02^{\rm h} 09^{\rm m} 37.38^{\rm s} \pm 0.03^{\rm s} $  & $+57\degr{} 59' 45.35'' \pm 0.26'' $& $55.3(6)$  & $2.09^{+0.21}_{-0.23}$ & $1.0639060415(1)$ & $20(2)$ & $21.4(5)$  \\
PSR J0854+5449 &  $08^{\rm h} 54^{\rm m} 25.733^{\rm s} \pm 0.002^{\rm s}$ & $+54\degr{} 49' 28.81'' \pm 0.01'' $& $18.837(1)$  & $0.75^{+0.15}_{-0.14}$  & $1.233032602667(5)$ & $9.3(0.1)$  & $21.5(6)$  \\
PSR J1252+53   & $12^{\rm h} 52^{\rm m} \pm 13^{\rm m}$ & $+53\degr{} 42' \pm 17'$  & $20.70(3)$   & $1.00_{-0.23}^{+0.33}$ & $0.22010358290(8)$ & $20(3)$  & $0.09(4)$ \\
PSR J1838+50   & $18^{\rm h} 38^{\rm m} \pm 8^{\rm m}$ & $+50\degr 51' \pm 15'$  & $21.81(1)$   & $1.54_{-0.17}^{+0.19}$ & $2.577223412(5)$ & $13(2)$ & $3.9(2)$  \\
PSR J1931+42   & $19^{\rm h} 31^{\rm m} \pm 7^{\rm m} $  & $+42\degr{} 30' \pm 5' $  & $50.90(2)$   & $3.13_{-0.40}^{+0.44}$ & -- & $32(6)$ & $8(1)$ \\
PSR J2108+4516 & $21^{\rm h} 08^{\rm m} \pm 7^{\rm m}$ & $+45\degr{}16'\pm 4' $& $82.4(3)$  & $3.37_{-0.43}^{+0.37}$ & $0.57722824(7)$ & $15.1(1)$ & $204(2)$ \\ 
\hline
\end{tabular}
\tablenotetext{a}{Based on \chimepsr{} single pulse detections unless otherwise stated in the text. Uncertainties are substantial underestimates for all sources using single pulse TOAs.}
\tablenotetext{b}{The full-width-at-half-maximum of a Gaussian fit to the profile (or the median of all the values for each single pulse detected).}
\tablenotemark{c}{Based on \chimefrb{} detections. The burst rate is calculated as the total number of detections with S/N $>8$ divided by the total time \chimefrb{} has observed the candidate position, assuming with Poissonian uncertainty for burst numbers.}
\end{sidewaystable}

\begin{table}[htbp]
\centering
\caption{Flux density measurements for all sources. \deleted{The range of peak flux densities for the intermittent sources are given in the $S^{\rm peak}_{600}$ column, while for the more persistent source we provide mean flux density estimate in the $S^{\rm mean}_{600}$ column.}
\label{flux_table}}
\begin{tabular}{lcc}
\hline\hline
Pulsar name  & $S^{\rm peak}_{600}$ (mJy) & $S^{\rm mean}_{600}$ (mJy) \\
\hline
PSR J0121+53    & 50--350  & --\\
PSR J0209+5759  & --      & $\gtrsim 0.3$\\
PSR J0854+5449  & --      & $\gtrsim 0.5$\\
PSR J1252+53    & 100--230 & --\\
PSR J1838+50   & 75--330  & --\\
PSR J1931+42 & 25--150 & --\\
PSR J2108+4516   &  --      & $\gtrsim 3$\\ 
\hline
\end{tabular}
\tablecomments{The flux density measurements here are lower limits. See Section~\ref{section:flux_density} for details.}
\end{table}

In Figure \ref{fig:shi_waterfalls}, we present dynamic spectra or ``waterfall'' plots and single pulse profiles from the \chimepsr{} system for sources detected only in single pulses. For sources with complete pulsar timing solutions, we present timing residuals in Figure~\ref{fig:timing_residuals} and single observation profiles in Figure~\ref{fig:persistent_profiles}.

\subsection{Source without spin period measurement: PSR J1931+42 \label{section:noperiod}}
Though we successfully measure a spin period for six of our seven sources, we have not yet done so for PSR J1931+42.

This candidate was first detected by \chimefrb{} on 2018-08-03 (MJD 58333) with a DM of 50.9\,\dmunits{}.
Search-mode \chimepsr{} follow-up was conducted, amounting to 40 observations spanning MJDs 58443--58542. 
PSR J1931+42 has only been detected in single pulses, generally with only one pulse per 20-minute observation.
An example pulse is shown in Figure~\ref{fig:shi_waterfalls}d.
\chimepsr{} detected a single pulse from this candidate on three separate days (MJDs 58534, 58537, and 58540), and detected two pulses separated by $\sim 380$ seconds on MJD 58523.
While there are more detections from \chimefrb{} (59 pulses over MJD 58372--58993), there are no observations with more than two detections, and only seven instances where two pulses were detected.
The smallest time difference between any two pulses detected with \chimefrb{} is $\sim 98$ seconds.

\subsection{Highly intermittent sources \label{section:SHI}}
Several of the sources reported have measured spin periods but display high intermittency. These sources may be RRATs or may be highly-intermittent pulsars, but do not have full timing solutions. Each period is initially determined via brute force methods using single pulse detections, then refined with pulsar timing software as discussed in \ref{section:timing_procedure}.

Figure \ref{fig:cf_detections_histograms} demonstrates that many of our sources are detected at low S/N, near our FRB detection cutoff. It is possible that some portion of our source's intermittency  may not be intrinsic but derived from failure to detect lower luminosity emission from these sources. Additionally, our relatively large position uncertainties restrict our abilty to optimally detect single pulses. These factors further muddy the RRAT vs. intermittent pulsar distinction, so we classify sources by their apparent intermittency and do not speculate as to their exact nature.

\subsubsection{PSR J0121+53}
\begin{figure*}[htbp]
\gridline{
    \leftfig{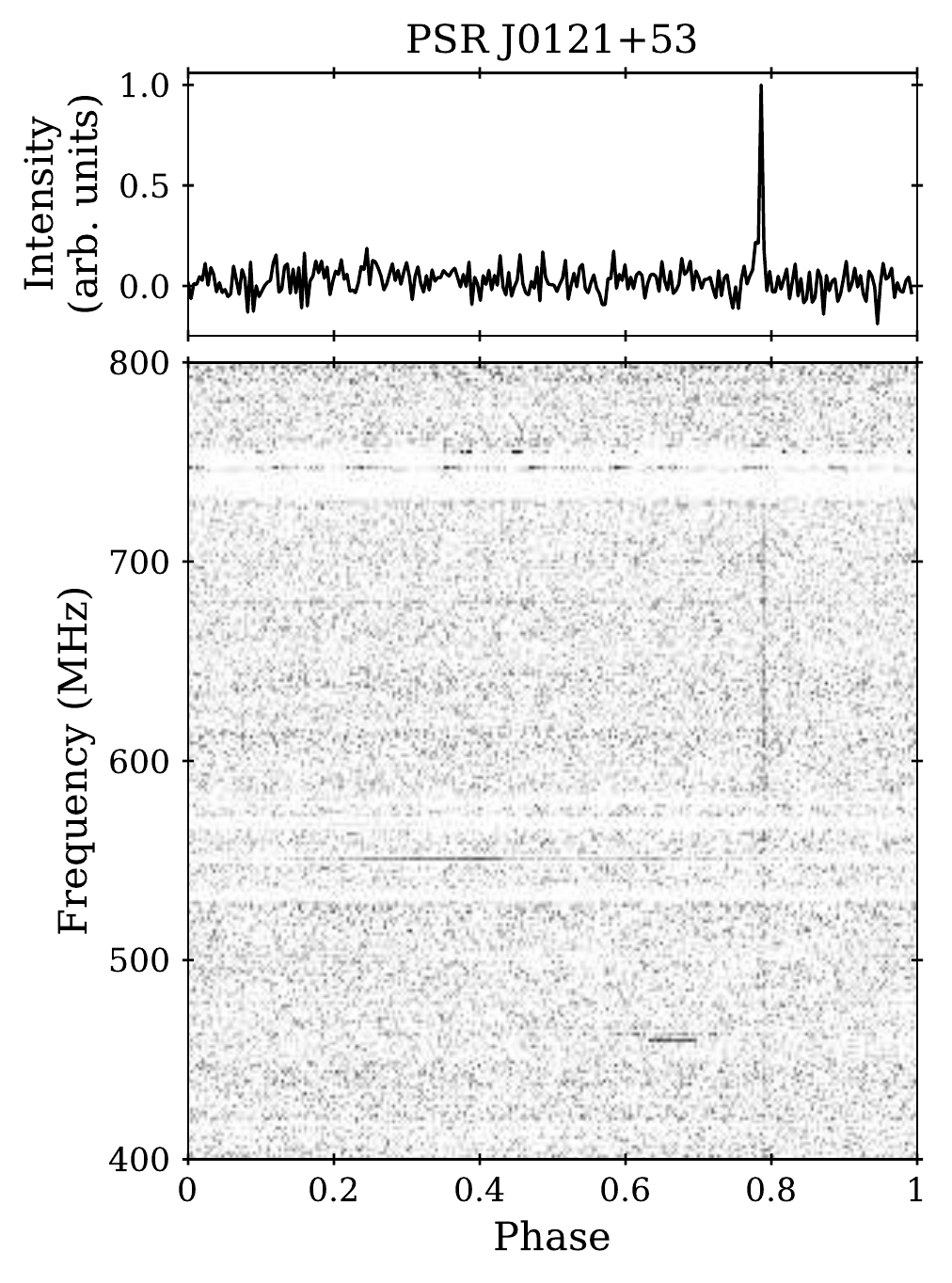}{0.4\textwidth}{(a)}
    \rightfig{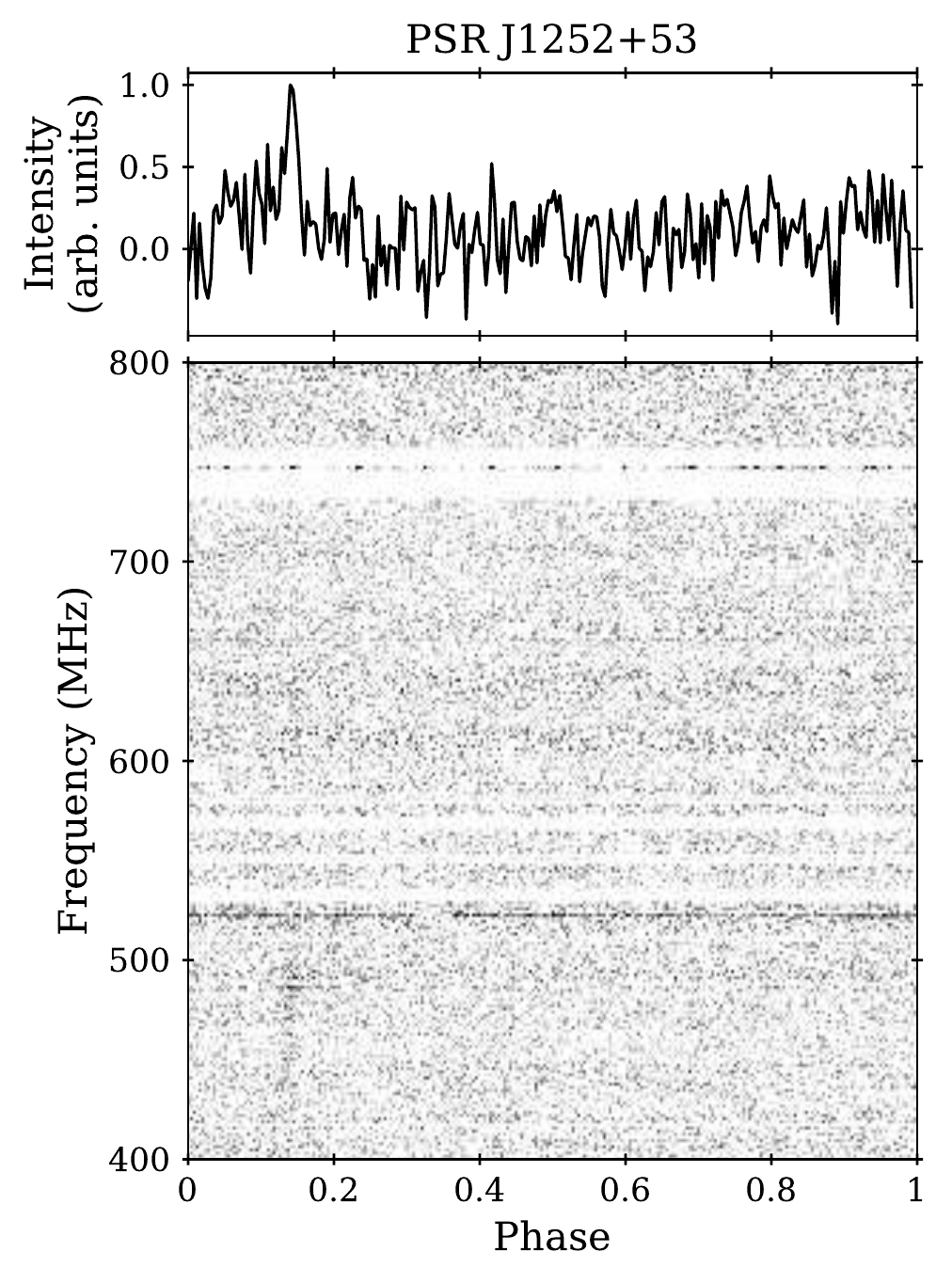}{0.4\textwidth}{(b)}
}
\gridline{
    \leftfig{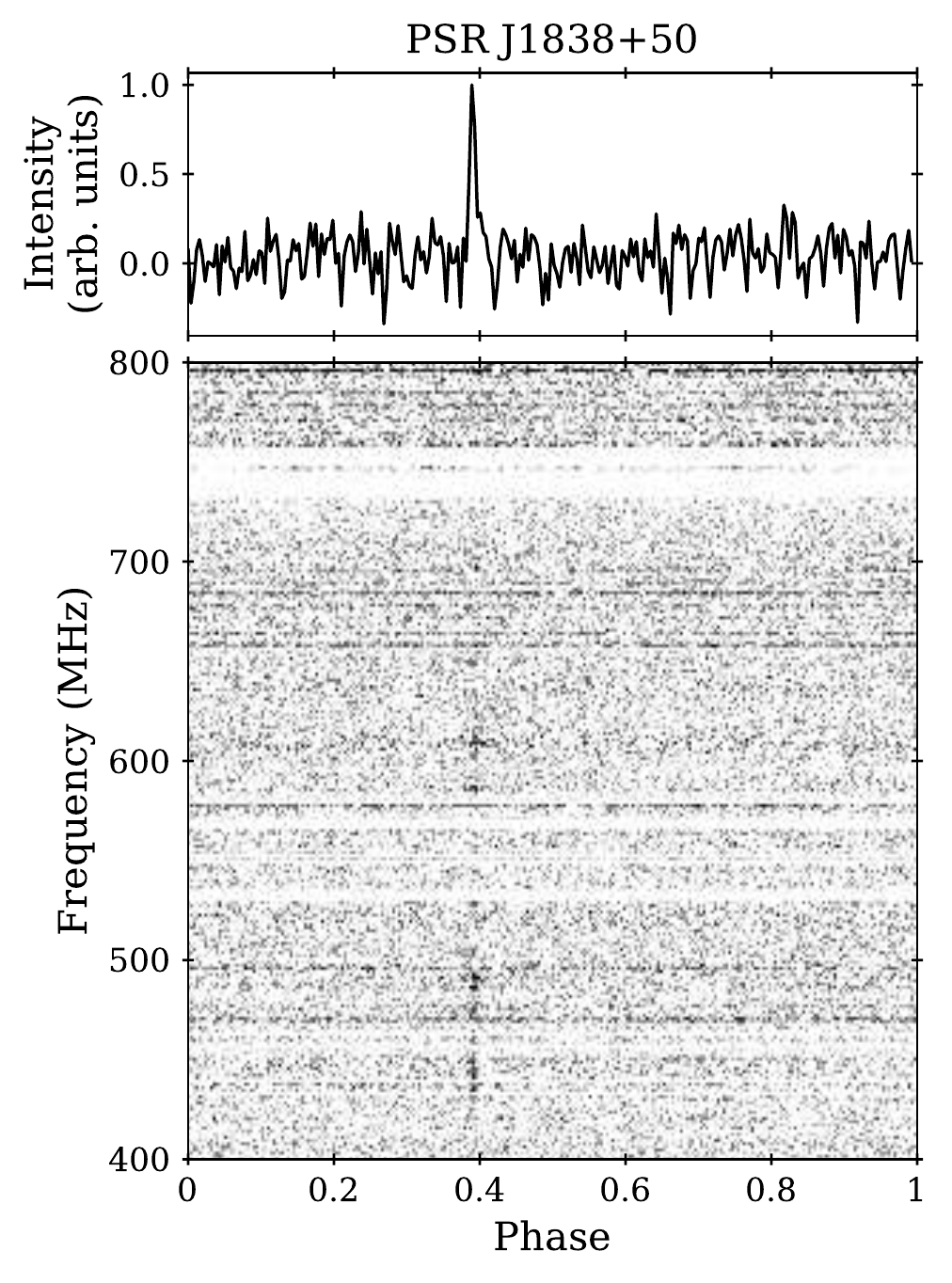}{0.4\textwidth}{(c)}
    \rightfig{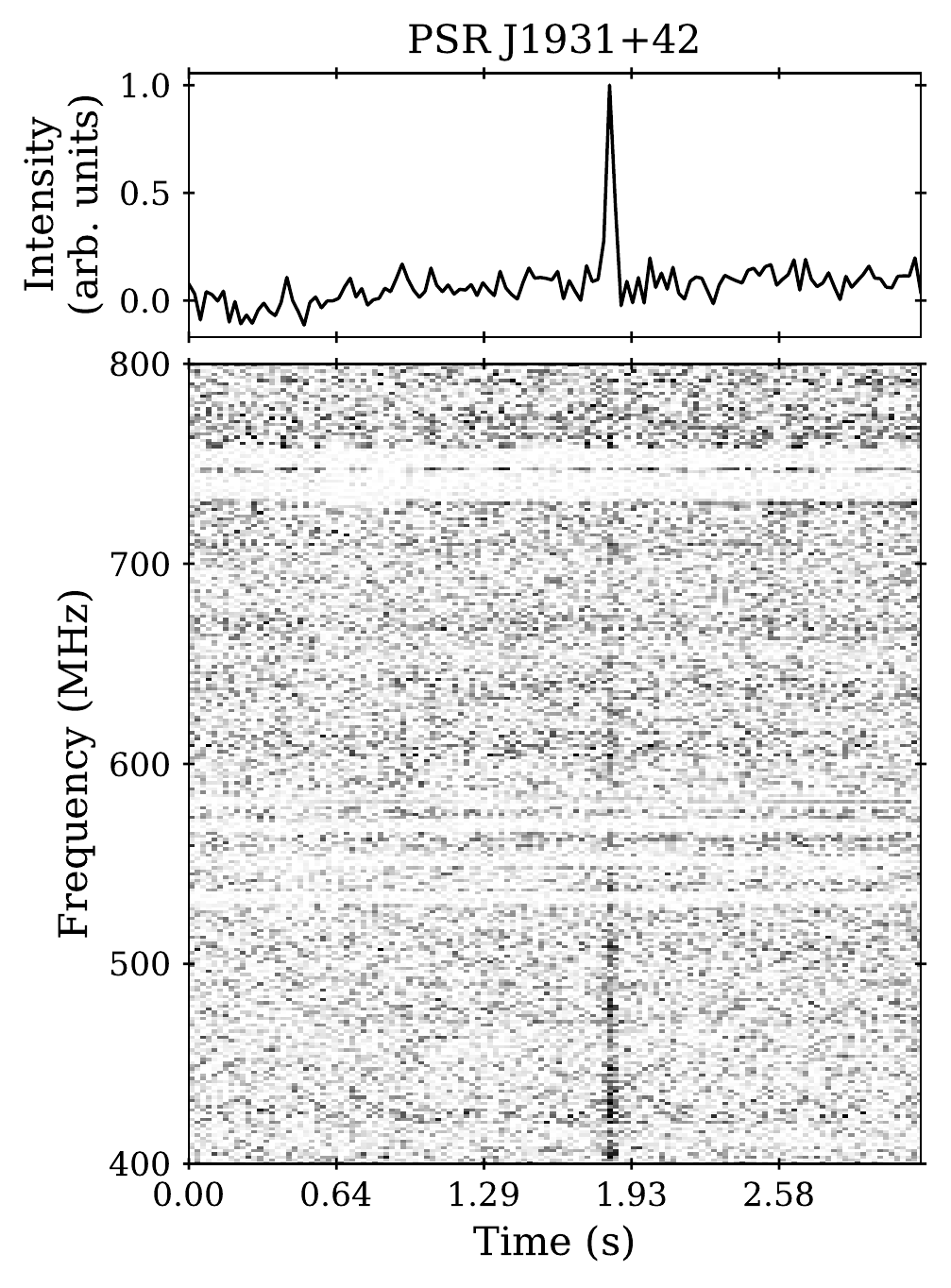}{0.4\textwidth}{(d)}
}
\caption{Single pulses detected by \chimepsr{} from each highly intermittent source. The single-pulse profile versus pulse phase is in the upper panels and the frequency spectrum versus pulse phase in the lower panels. (a) J0121+53 detected on 2020-03-04 (MJD 58912) with a peak S/N $\approx$ \replaced{14}{22}, corresponding to a peak flux density $\approx 281$\,mJy; (b) J1252+53 detected on 2020-03-10 (MJD 58918) with a peak S/N $\approx$ \replaced{8}{5}, corresponding to a peak flux density $\approx 190$\,mJy; (c) J1838+50 detected 2019-03-01 (MJD 58543) with a peak S/N $\approx$ \replaced{14}{9}, corresponding to a peak flux density $\approx 334$\,mJy; and (d) J1931+42 detected on 2019-02-09 (MJD 58523) with a peak S/N $\approx$ \replaced{19}{22}, corresponding to a peak flux density \replaced{$\approx 121$\,mJy}{$\approx 146$\,mJy}, where the horizontal axis for this source is in units of time since we do not have a period estimate for this source. \label{fig:shi_waterfalls}}
\end{figure*}

PSR J0121+53 was first detected by \chimefrb{} on 2018-09-05 (MJD 58366) at a DM of 87.4\, \dmunits. Single pulse detections with \chimepsr{} search-mode data indicate the DM is slightly higher: 91.38\, \dmunits; \replaced{Though this is a large discrepancy, such uncertainties are possible in initial detections due to \chimefrb{}'s tree dedispersion structure. For this reason, final DM calculations in \chimefrb{} publications such as \cite{chimefrb_repeaters19} and \cite{fab+20} do not publish initial DM determinations, but instead fit for a structure maximizing DM.}{the discrepancy arises due to large uncertainties in \chimefrb{} DM measurements from the real-time pipeline.}
The source is seen consistently in single pulse plots, but has not been successfully folded. 
\deleted{This source has the highest DM of the seven sources presented here, but since it is in the Galactic Plane it is at a distance of roughly 3.3 kpc \citep{cl02}. An example pulse is show in Figure~\ref{fig:shi_waterfalls}a. }
Though this pulse is single peaked, some pulses from this source have shown double-peaked structure, indicating \replaced{mode changing or subpulse drifting}{potentially interesting morphological characteristics.} 

\replaced{Analysis with the new multi-day RRAT period finder determines an underlying spin period of 2.725 s. We used \chimefrb{} metadata detections to construct an initial solution, which found a spin period of 2.726 s. }{The spin period was initially estimated using the multi-day RRAT period finder and \chimefrb{} metadata.} The final value is reported in Table \ref{tab:parameter_table} along with other parameters for this source.

\subsubsection{PSR J1252+53}
PSR J1252+53 was first detected by \chimefrb{} on 2018-10-23 (MJD 58414) at a DM of 21.03\,\dmunits, \replaced{and subsequently search mode observations were conducted using \chimepsr{}, which determined the DM to be}{though \chimepsr{} observations found a DM of} 20.70\,\dmunits. PSR J1252+53 appears only four times in over 100 near-daily observing sessions. Results from \chimepsr{} single pulse detections indicate that the underlying period is 220 ms; no \chimefrb{} metadata solution has been found. 

Although detected, it is clear \replaced{in Figure \ref{fig:shi_waterfalls}b}{from \chimefrb{} detections in Figure \ref{fig:cf_detections_histograms}} that PSR J1252+53 has a lower S/N than our other six sources. It is also the source with the fewest \chimefrb{} and \chimepsr{} detections.

\subsubsection{PSR J1838+50}
PSR J1838+50 was first detected by \chimefrb{} on 2018-07-27 (MJD 58326) with a DM of 21.8\,\dmunits{}. PSR J1838+50 was followed up regularly using \chimepsr{} search mode observations between MJD 58463 and 58561 and again between MJD 59033 and 59123. \chimefrb{} metadata analysis and \chimepsr{} single pulse analysis independently find a period of 2.577\,s. An example pulse is show in Figure~\ref{fig:shi_waterfalls}c. Like PSR J0121+53, PSR J1838+50 is sufficiently intermittent that folding data for the duration of an observation does not improve the detection.
\deleted{Due to scheduling and data storage constraints, PSR J1838+50 has the fewest observations of the seven sources presented here. However, the \chimefrb{} burst rate suggests this source is moderately active. This raises the possibility that the solution could be substantially improved by continued observation. }

\subsection{Persistent Sources \label{section:persistent_sources}}
The other sources presented here are all likely ``slow'' pulsars. \chimefrb{} is unlikely to detect millisecond pulsars due to its intrinsic time resolution (0.983 ms). These sources demonstrate some periods of inactivity or more limited activity, providing a possible explanation for the failure to detect them in previous pulsar searches.

\begin{figure*}[hbtp]
    \centering
    \includegraphics[width=0.48\textwidth]{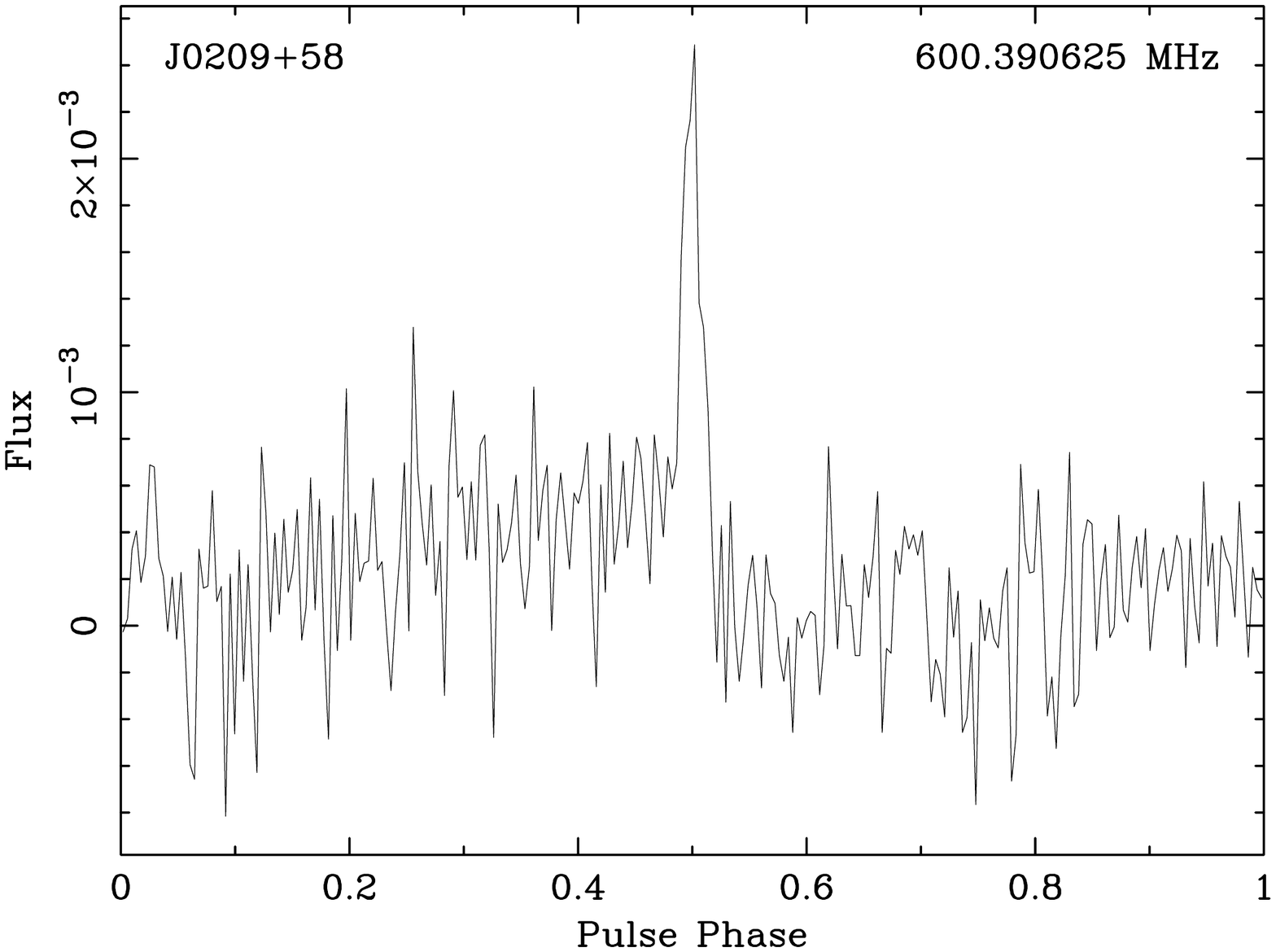}
    \includegraphics[width=0.48 \textwidth]{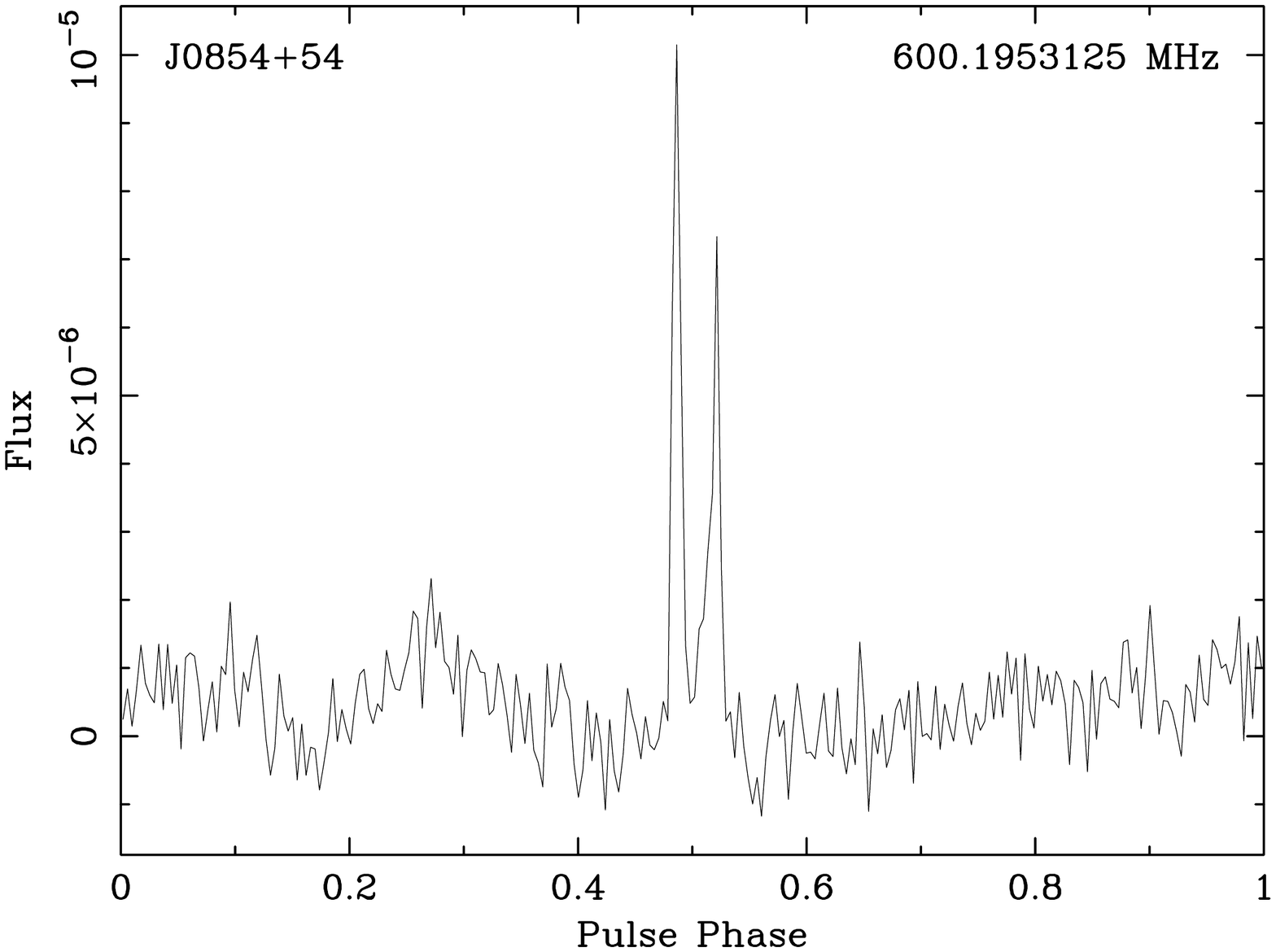}
    \includegraphics[width=0.48\textwidth]{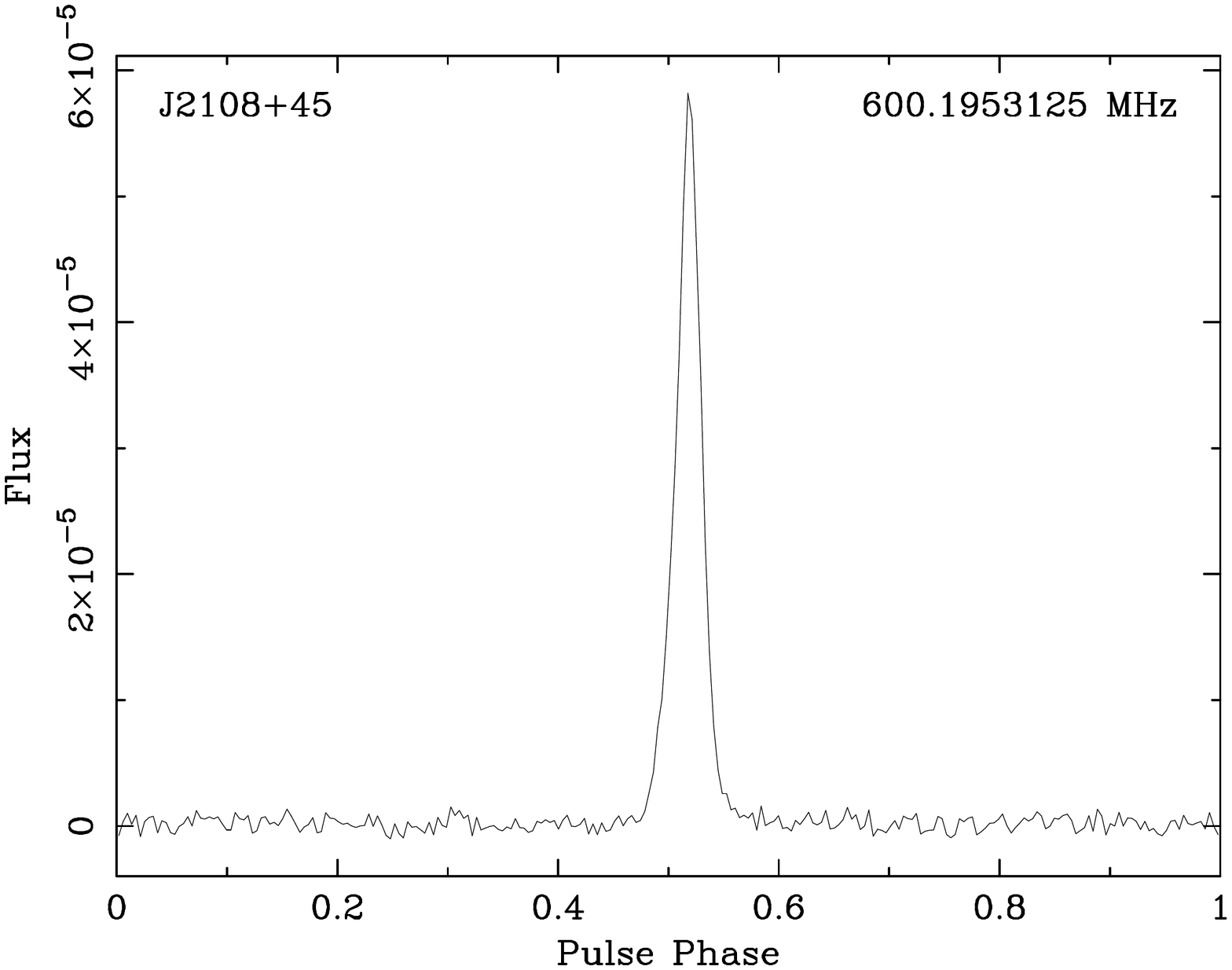}    
    \caption{Single observations profiles for J0209+5759, J0854+5449, and J2108+4516. 
    J0209+5759 has S/N 11.6, J0854+5449 has S/N 21.5, and J2108+4516 has S/N 196.}
    \label{fig:persistent_profiles}
\end{figure*}

For these sources, we are able to determine complete pulsar timing solutions. We present complete solutions and timing residuals for isolated pulsars PSR J0209+5759 and PSR J0854+5449 in Table \ref{tab:timing_solutions} and Figure \ref{fig:timing_residuals}. We show pulse profiles for single observations for each source in Figure \ref{fig:persistent_profiles}. We choose to present single observation profiles instead of average profiles as these sources are intermittent. Due to its greater complexity, the complete solution for PSR J2108+4516 is deferred to a future publication. 

\subsubsection{PSR J0209+5759}
PSR J0209+5759 was initially detected by \chimefrb{} on 2018-09-06 (MJD 58367) with a DM of 56.6 \dmunits{}, but subsequent PRESTO analysis demonstrated the actual DM was slightly lower (55.3 \dmunits{}). This source was observed in search-mode by \chimepsr{} on 293 days between MJD 58411 to 58922. Since MJD 59100, we have been taking daily fold-mode observations of this source. PSR J0209+5759 displays a high degree of nulling, with 1,500--1,800 second observations generally including fewer than five single pulse detections. After folding, PSR J0209+5759 continues to display intermittent emission within observations. We did not detect a rotation measure for this source, using the method outlined in \cite{npn+20}. Analysis following the procedure outlined in \cite{nwm+20} finds a lower limit nulling fraction of 21\%. However, we are able to determine a full timing solution for this source, with measured and derived parameters presented in Table~\ref{tab:timing_solutions}. Timing residuals are shown in the first panel of Figure~\ref{fig:timing_residuals}.

\subsubsection{PSR J0854+5449}
\begin{figure*}[htbp]
    \centering
    \includegraphics[width=0.48\textwidth]{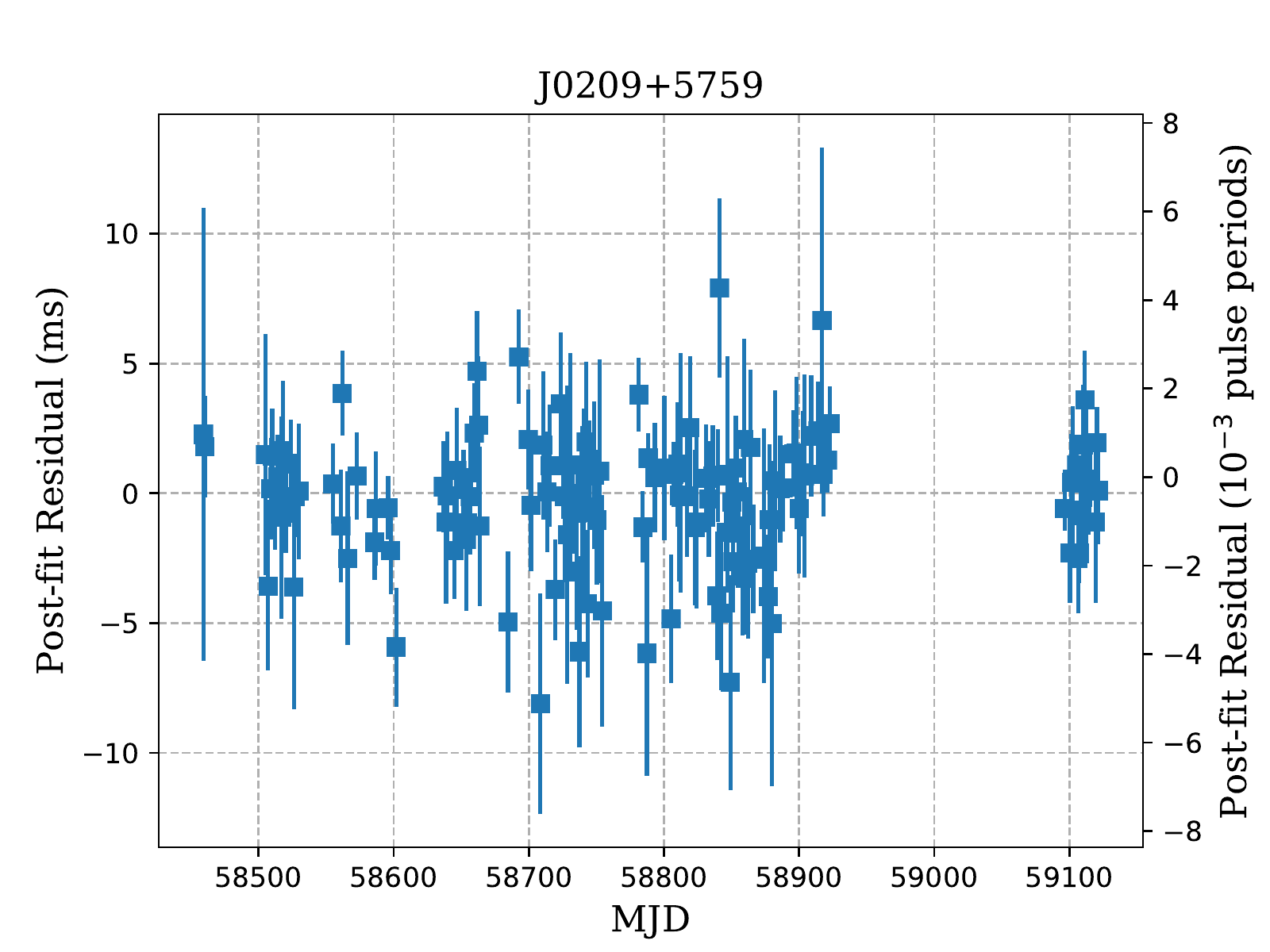}
    \includegraphics[width=0.48\textwidth]{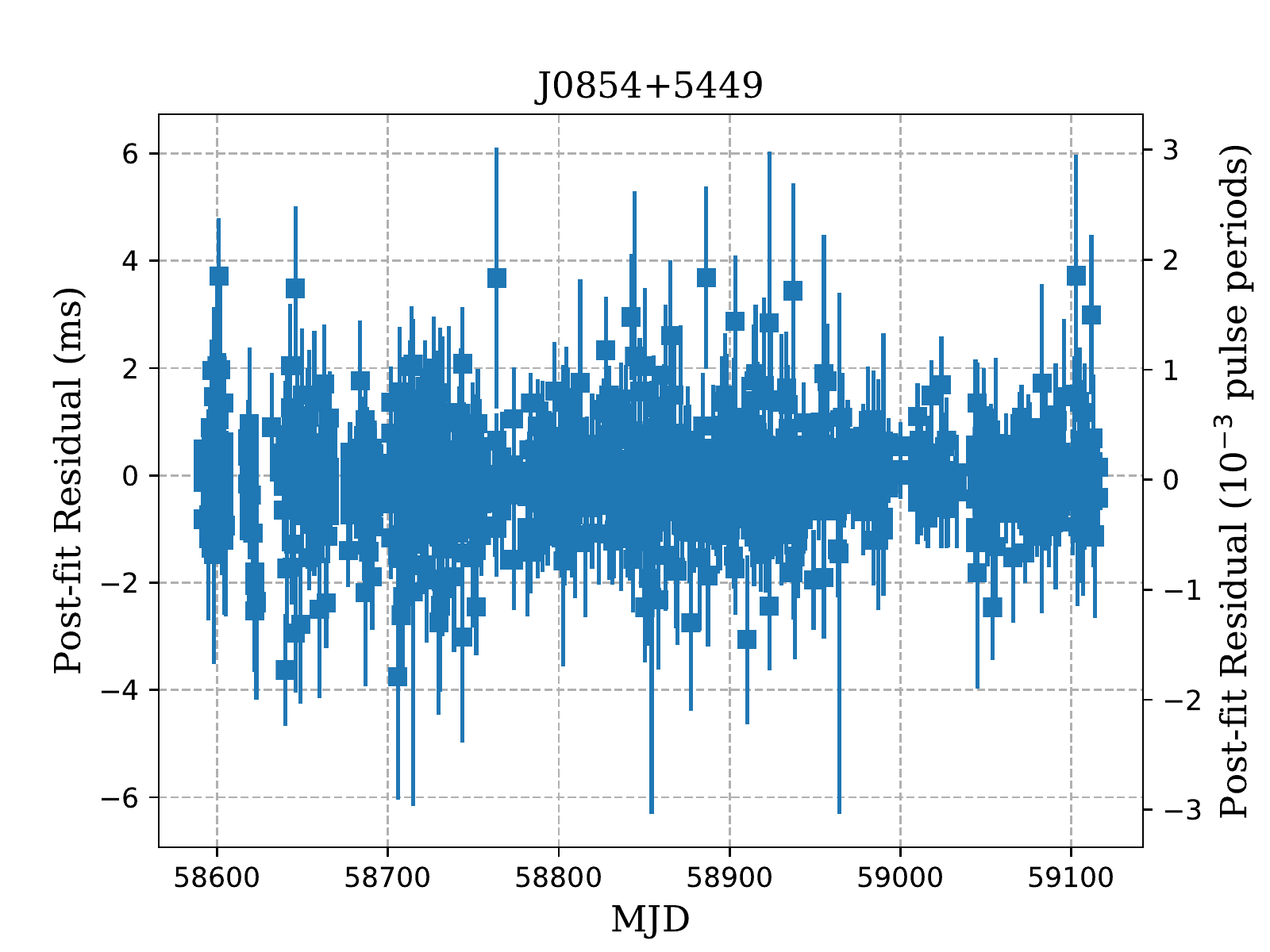}
    \caption{Best-fit timing residuals of PSRs J0209+5759 and J0854+5449 estimated from CHIME/Pulsar filterbank and fold-mode data. The smaller data set for J0209+5759 is due to fewer pulses emitted by the source, in a manner consistent with other known RRATs. Best-fit parameters of the timing model for these sources are shown in Table \ref{tab:timing_solutions}. \label{fig:timing_residuals}}
\end{figure*}

PSR J0854+5449 was first detected by \chimefrb{} on 2018-07-25 (MJD 58324) with a realtime-detected DM of 17.8 pc cm$^{-3}$. We began near-daily filterbank observations of this position with \chimepsr{} on MJD 58592, and switched to coherent dedispersion fold-mode observations on MJD 58868 once an adequate timing solution was derived. We derived a coherent timing solution from the combined filterbank and fold-mode data sets with parameters that are consistent with implying the source is an isolated, slow pulsar. J0854+5449 has a rotation measure value of $-8.8 \pm 2$ rad m$^{-2}$, determined using the procedure described in \cite{npn+20}. Measured and derived parameters are given in Table~\ref{tab:timing_solutions} and the timing residuals are shown in the second panel of Figure~\ref{fig:timing_residuals}.

\subsubsection{PSR J2108+4516}
First detected by \chimefrb{} on 2018-10-11 (MJD 58402) and subsequently monitored with regular \chimepsr{} filterbank and fold-mode observations from MJD 58534 onward, PSR J2108+4516 is a complex binary system that undergoes periods of substantial eclipse and significant DM and scattering variations. A detailed study of this source, including a full timing solution and a demonstration of its DM and scattering properties, will be presented in an upcoming paper (Andersen et al., in prep.).

\begin{table*}
\centering
\caption{Best-fit parameters and derived quantities for PSRs J0854+5449 and J0209+5759. \label{tab:timing_solutions}}
\begin{tabular}{p{65mm}ll}
\hline\hline
\multicolumn{3}{c}{Global Parameters \& Best-fit Metrics} \\
\hline
Pulsar name\dotfill & J0854+5449 & J0209+5759\\
Reference epoch (MJD)\dotfill & 58854.0 & 58790.0  \\
Observing timespan (MJD)\dotfill &  58592--59115 &  58459--59121\\
Number of Sub-bands\dotfill & 32 & 1 \\
Reduced $\chi^2$\dotfill & 1.04 & 1.03 \\
RMS timing residual (\us)\dotfill & 338 & 1960 \\
TOA uncertainty scale factor\tablenotemark{a}\dotfill & 1.1 & 1.6 \\
Ephemeris \dotfill & DE436 & DE436 \\
Clock Standard \dotfill & TT(BIPM2017) & TT(BIPM2017) \\
\hline
\multicolumn{3}{c}{Timing Solutions} \\
\hline
Right ascension (J2000), $\alpha$\dotfill &  08$^{\rm h}$54$^{\rm m}$25.7255(8)$^{\rm s}$ & 02$^{\rm h}$09$^{\rm m}$37.304(16)$^{\rm s}$\\
Declination (J2000), $\delta$\dotfill &  +54\degr 49\arcmin 28.782(11)\arcsec & +57\degr 59\arcmin 45.31(18)\arcsec \\
Pulse frequency, $\nu$ (s$^{-1}$)\dotfill & 0.8110085638790(6) & 0.939932413620(12) \\
Frequency derivative, $\dot{\nu}$ (s$^{-2}$)\dotfill & $-2.0535(14)\times 10^{-16}$ & $-1.22971(12) \times 10^{-14}$ \\
Dispersion measure, DM (\dmunits)\dotfill & 18.8368(15) & 55.282\tablenotemark{b}  \\
\hline
\multicolumn{3}{c}{Derived Quantities} \\
\hline
DM distance from NE2001 (kpc) \dotfill & 0.75$^{0.15}_{-0.14}$ & 2.09$^{+0.21}_{-0.23}$ \\
Galactic latitude, $l$ (deg)\dotfill & 162.8 & 133.2 \\
Galactic longitude, $b$ (deg)\dotfill & 39.4 & $-3.3$ \\
Characteristic age, $\log_{10}$($\tau_{\mathrm{c}}$ (yr))\dotfill & 7.79 & 6.15 \\
Surface magnetic field, $\log_{10}$($B_{\mathrm{surf}}$ (G))\dotfill & 11.8 & 12.5 \\
Nulling Fraction \dotfill & -- & 0.21 \\
Rotation Measure, rad m$^{-2}$ \dotfill & $-$8.8(20) & -- \\
\hline                                                                                                
\end{tabular}%
\tablenotetext{a}{This scale factor is multiplied to all raw TOA uncertainties in order adjust the reduced $\chi^2$ to the value stated in Table \ref{tab:timing_solutions}.}
\tablenotetext{b}{This DM value was estimated using the PRESTO {\tt prepfold} tool's DM search feature.}
\tablecomments{Measured parameter uncertainties from \tempo{} reflect 1-$\sigma$ confidence intervals, are denoted above in parentheses and apply to the last significant digit(s).}
\end{table*}

\subsection{Comparison to known RRATs}
Comparing the distribution of known RRAT properties to those presented here is not straightforward, given that the sample is both small and \replaced{non-normally}{not normally} distributed. This small sample also reduces the robustness of a chosen statistic estimation and often leads to an uncertain interpretation. Moreover, there is no accounting of selection bias in either sample. Under the above caveats, we present a rudimentary comparison, but reserve a more rigorous study of the \replaced{candidate}{source} property distributions to a later paper, when the CHIME sample size is larger.

\begin{figure*}[htbp]
\centering
\includegraphics[width=0.48\textwidth]{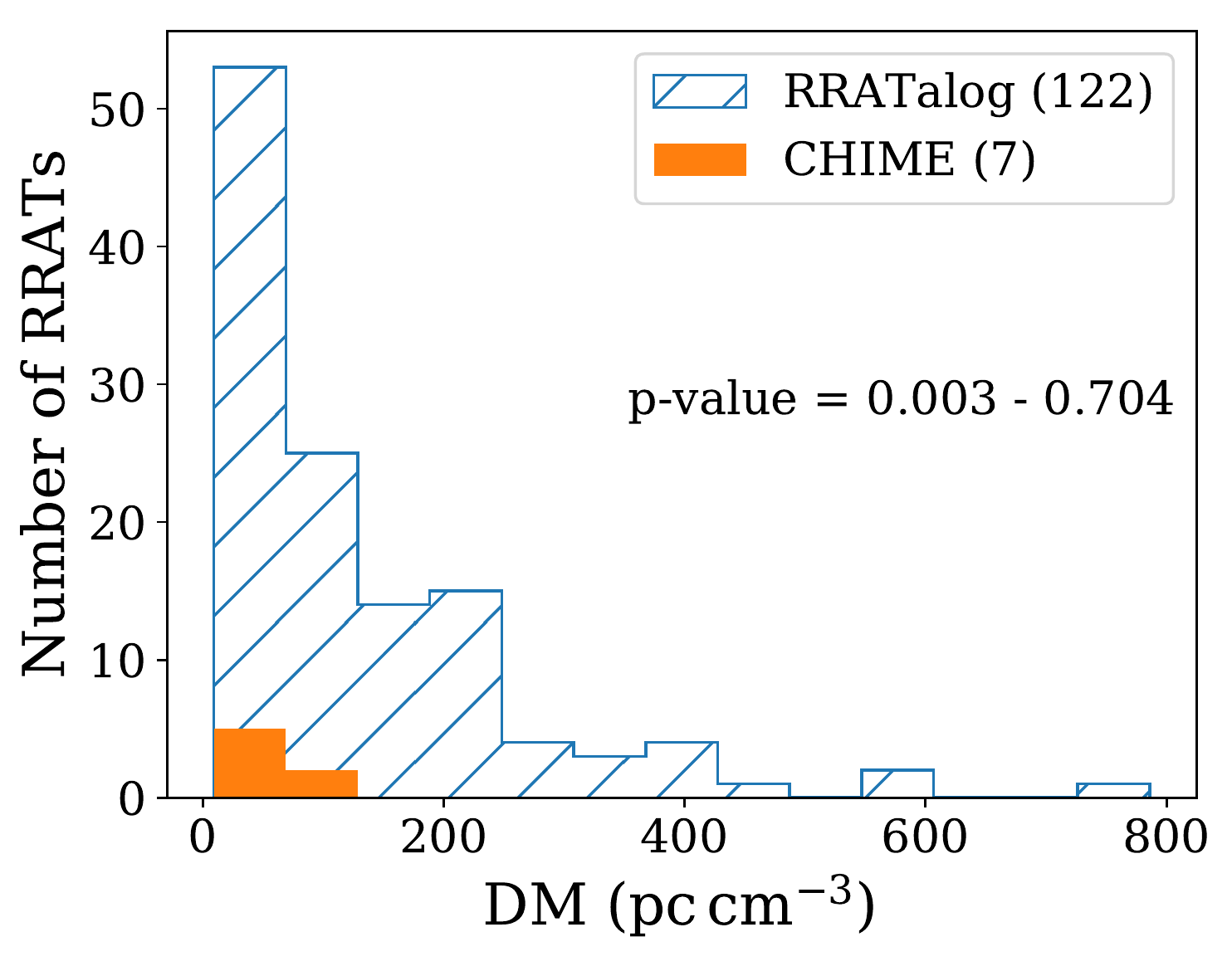}
\includegraphics[width=0.48\textwidth]{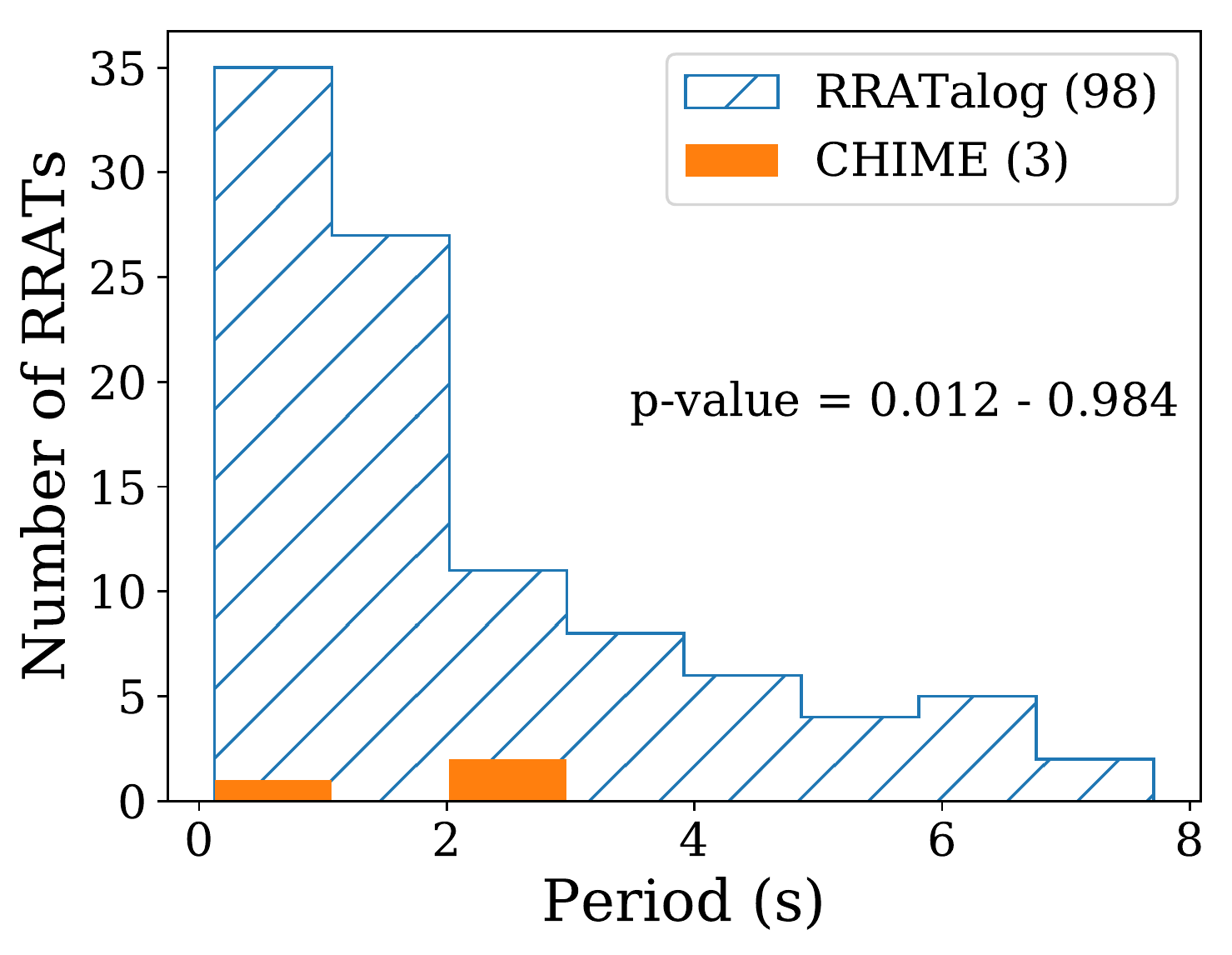}
\includegraphics[width=0.48\textwidth]{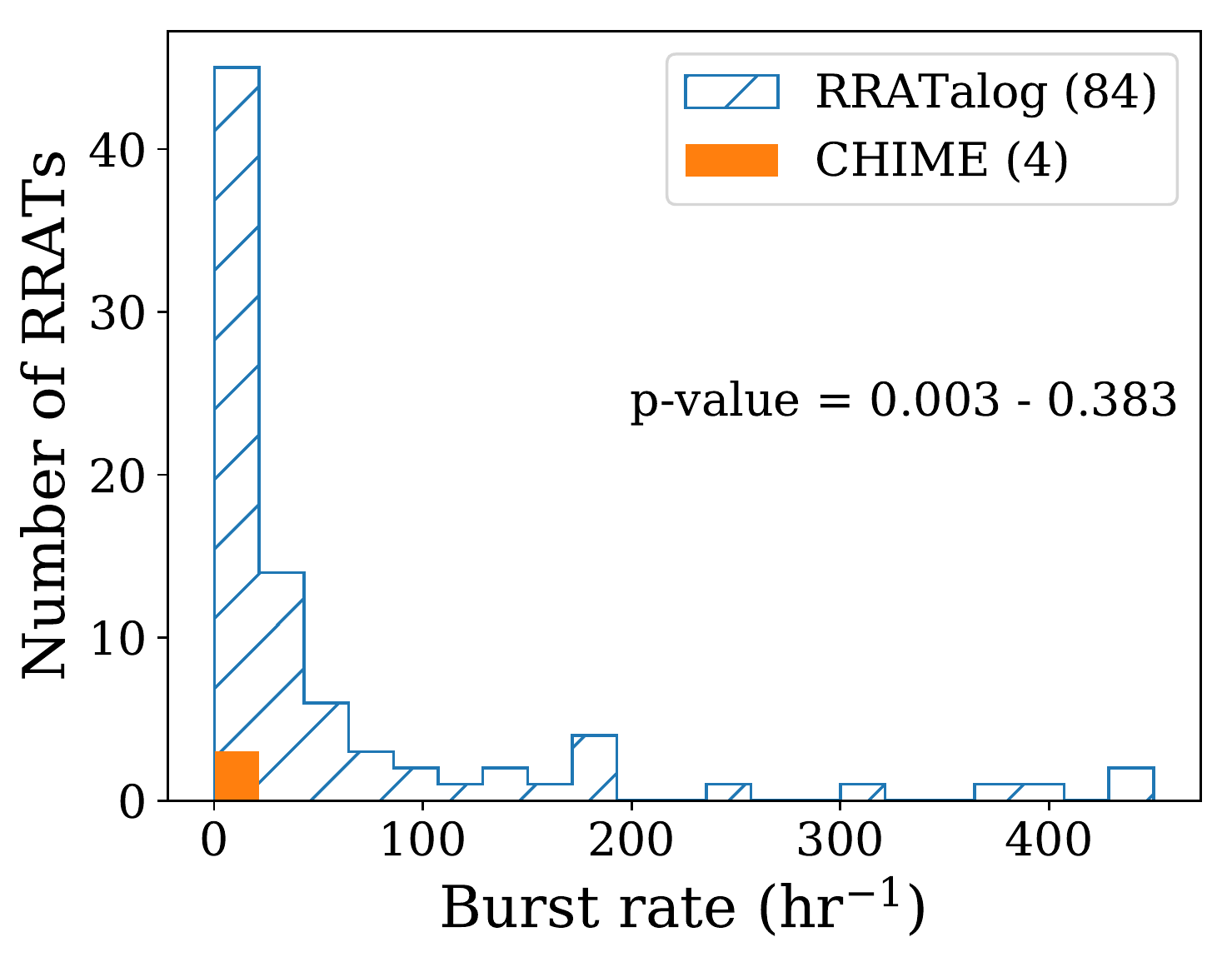}
\includegraphics[width=0.48\textwidth]{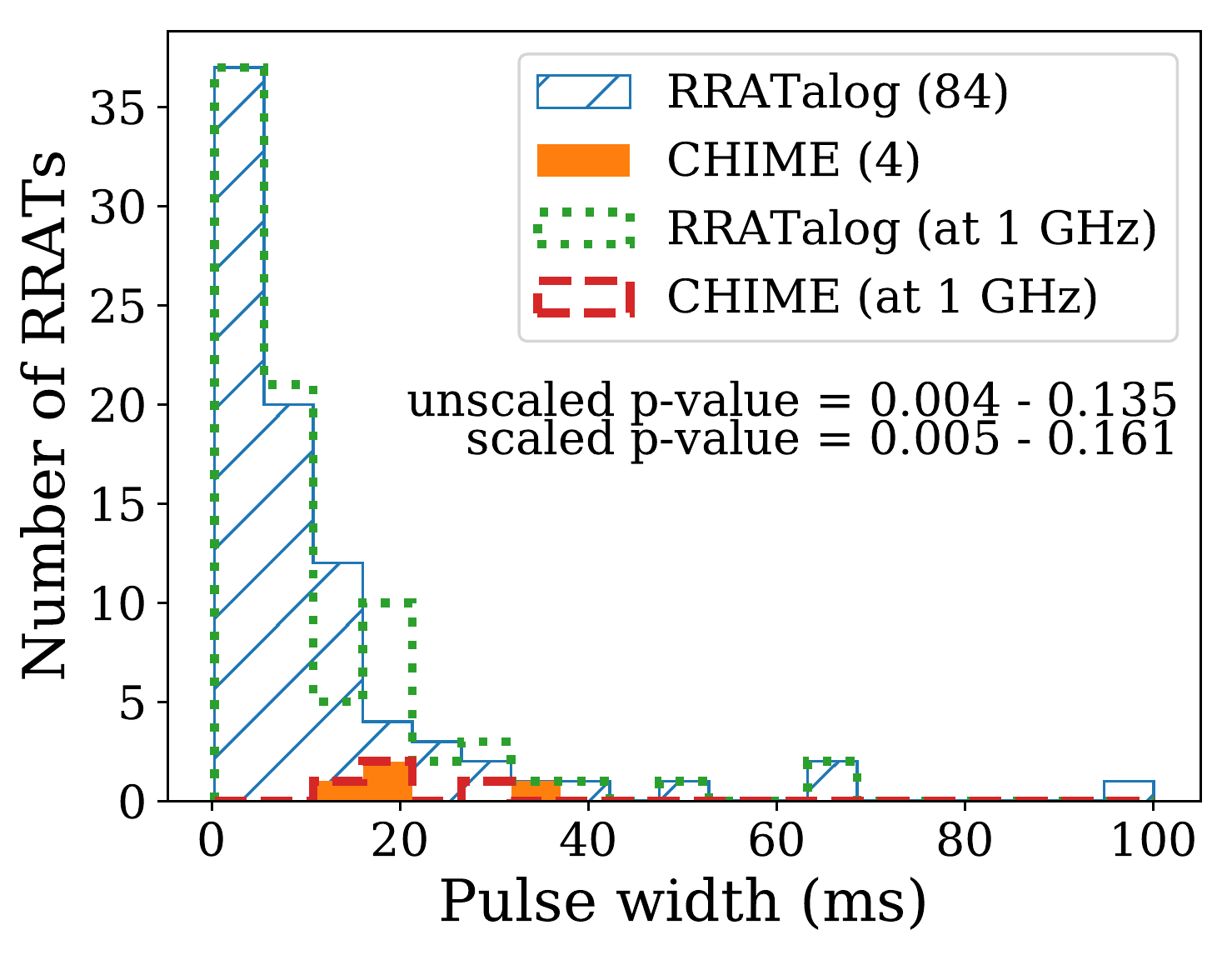}
\caption{Dispersion measure (top left), rotation period (top right), burst rate (bottom left), and pulse width (bottom right) distributions of known RRATs (blue histogram) and the CHIME candidates presented here (orange histogram). For the width distributions, the scaled (to 1 GHz) and unscaled versions are overlaid (see text for details). Samples sizes are given in parentheses in the legend keys. The range of p-values from the bootstrapped Kruskal-Willis H-tests are also annotated (and can be found in Table~\ref{tab:sample_compare}. For all parameters tested we cannot reject the null hypothesis that the samples are drawn from an underlying distribution with the same shape and median values. \label{fig:rrat_dists}}
\end{figure*}

The DM, period, burst rate, and pulse width distributions for both known RRATs and CHIME \replaced{candidates}{sources} are shown in Figure~\ref{fig:rrat_dists}.
\added{We exclude PSRs J0209+5759, J0854+5449 and J2108+4516 from the analysis because, even though they were initially detected by their single-pulse emission, they exhibit persistent emission.}
To compare the parameters distributions, we employ the \replaced{Mann-Whitney $U$-test}{ Kruskal-Willis H-test}\footnote{Specifically, the \replaced{{\tt scipy.stats.mannwhitneyu}}{{\tt scipy.stats.kruskal}} implementation. We note that, since there are $\gtrsim 25$ members of the known RRAT group, asymptotic approximations are made by the method when computing the p-values (i.e., they are not exact).}. \added{If we assume that the distribution shapes are the same between groups (i.e., between the CHIME sample and the known RRAT sample), then the null hypothesis is that both distribution medians are equal.}
Given the small sample size we also bootstrap the statistic computation, resampling both the CHIME and known RRAT samples with replacement and iterating $10^4$ times for each distribution. 
The confidence intervals of the p-values from these iterations (see Table~\ref{tab:sample_compare}) are then compared to the nominal threshold in order to decide whether the null hypothesis can be rejected. 
Specifically, we reject the null hypothesis only if the upper confidence bounds are below 0.05.
Under that criterion, for all parameters compared, we cannot reject the hypothesis that both samples are drawn from the same distribution.

\begin{table}[!htbp]
\centering
\caption{Confidence intervals of p-values from the bootstrapped Kruskal-Willis H-tests comparing the CHIME and known RRAT distributions. 
\label{tab:sample_compare}}
\begin{tabular}{lcc}
\hline\hline
Parameter       &  \multicolumn{2}{c}{95\% confidence interval}\\
                & Lower & Upper \\
\hline
DM              & 0.009 & 0.878 \\
Period          & 0.012 & 0.984 \\
Burst rate      & 0.003 & 0.383 \\
Pulse width     &    & \\
\dotfill Raw    & 0.004 & 0.135 \\
\dotfill Scaled & 0.005 & 0.161 \\
\hline
\end{tabular}
\end{table}

The pulse widths listed in Table~\ref{tab:parameter_table} and determined as described in Section~\ref{section:width} are referenced to 600\,MHz.
Widths from the literature are measured at various frequencies, thus it is perhaps unwise to compare the values directly.
If we invoke radius-to-frequency mapping, then a measured pulse width has some dependence on the observing frequency, $\nu$.
Specifically, \citet{kg98} suggest that $W_{50}\propto r_{\rm em}^{0.5}$, where $r_{\rm em} \propto \nu^{-0.2}$ is the emission altitude (i.e., $W_{50}\propto \nu^{-0.1}$).
Incorporating this idea and scaling all measured pulse widths to 1\,GHz, we recompute the distributions and apply the same statistical test as before. 
Again, we cannot reject the null hypothesis. 

\subsection{Potential for future discoveries \label{Section:population}}

\chimefrb{} was designed to be sensitive to bright single-pulse events, and is thus expected to be adept at detecting single pulses from pulsars and RRATs. 
However, the single pulse characteristics of pulsars are, in general, poorly understood. \added{For example, the single pulse luminosity distribution is not well understood for any single pulsar and for the population in general, although there have been attempts to make headway in this area \citep[e.g.,][]{Mickaliger2018}.}
The \replaced{single pulse characteristics of RRATs are even more poorly understood}{state of constraints on single pulse characteristics of RRATs is even worse} due to small number statistics (only \replaced{$\sim 100$}{$\sim 130$} confirmed detections)\footnote{See the RRATalog: \url{http://astro.phys.wvu.edu/rratalog/}.}, even though estimates of the number of RRAT-like objects in the Galaxy are similar to that of the ``normal'' pulsar population size ($\sim 10^5$; \citealp{mll+06}).
\chimefrb{} is unique in terms of instrument response functions, biases and selection effects. These instrumental factors are currently being investigated further by the injections pipeline,\added{ discussed briefly in \cite{cfrbcatalog} and to be discussed in detail in} Merryfield et al. (in prep). This makes determining the number of potential pulsar discoveries with \chimefrb{} challenging. Nevertheless, we present an initial estimate of the number of potential \chimefrb{} pulsar and RRAT discoveries.

\added{In our estimate, we assume only one population of pulsars, namely the ``regular'' pulsars, as it is not clear whether RRATs are an effect of telescope sensitivity or a unique set of sources with different intrinsic properties \citep{kkl+11}. Thus, the only difference between \chimefrb{} and other telescopes is that \chimefrb{} primarily operates in single pulse mode for pulsar finding, leading to a heavy dependence on the single pulse luminosity distributions. Given these assumptions we use the pulsar population synthesis code, \mbox{PSRPopPy} \citep{blr+14}, to simulate $N_{\rm sim} \sim 1.1\times10^5$ pulsars. We selected these numbers by continually simulating pulsars with \mbox{PSRPopPy} using the Parkes Multibeam Survey (PKMBS) parameters until we detected the total number of pulsars found by the survey \citep{Lyne1998}. This simulated population is therefore close to the Galactic pulsar estimate in \citep{Lyne1998,Lorimer2006}. The pulsars are also each assigned standard pulsar parameters like spectral index, period, and pulse width according to the default settings of \mbox{PsrPopPy}.}

\added{As the distribution of the single pulse luminosity is not well understood, we adopt a log normal model \citep{Mickaliger2018}, 
\begin{equation}
    p(L) = \frac{1}{L\sigma_L \sqrt{2\pi}} \exp{\bigg(-\frac{1}{2}\frac{(\ln{L}-\mu_L)^2}{\sigma_L^2}\bigg)}
\end{equation}
where $L$ is the drawn luminosity, $p(L)$ is the probability density, $\mu_L$ is the log mean luminosity and $\sigma_L$ is the scale parameter. The mean luminosity distribution is derived from \cite{fk06}, a \mbox{PSRPopPy} default.\\
However, to the best of our knowledge, the scale parameter $\sigma_L$ for the single pulse luminosity is not well constrained. Therefore, we calibrate the $\sigma_L$ parameter against the $278$ PKMBS single pulse detections in \cite{Mickaliger2018}, McLaughlin (priv. com.). Like how $\mu_L$ is drawn from a distribution, $\sigma_L$ for each pulsar needs to also be drawn from a distribution. From \cite{Mickaliger2018}, we find that the distribution of $\sigma_L$ is Gaussian. Thus, 2 parameters are calibrated for using the PKMBS, $\mu_{\sigma_L}$, the mean of $\sigma_L$, and $\sigma_{\sigma_L}$, the standard deviation of $\sigma_L$. $\sigma_L$ is drawn from 
\begin{equation}
    f(\sigma_L)=\frac{1}{\sigma_{\sigma_L}\sqrt{2\pi}}\exp{\bigg(-\frac{1}{2}\frac{(\sigma_L-\mu_{\sigma_L})^2}{\sigma_{\sigma_L}^2}\bigg)}
    \label{eq:single_pulse_lum}
\end{equation}
where $f(\sigma_L)$ is the probability density. As $\sigma_L$ can not be a negative value, we redraw from eq. \ref{eq:single_pulse_lum} if a negative value is drawn.
}
\added{First we calculate the number of pulses that a pulsar ought to have emitted given the pointing time of the PKMBS,
\begin{equation}
    N_{\rm pulse}=\frac{T_{\rm Parkes}}{P}
\end{equation}
where $N_{\rm pulse}$ is the number of pulses emitted by that particular pulsar, $T_{\rm Parkes}=30\,\text{minutes}$, is the pointing time and $P$ is the pulsar period. Then for each pulse we draw from the single pulse luminosity distribution and calculate whether the pulsar would have been detected by the PKMBS via the single pulse radiometer equation
\begin{equation}
    {\rm S/N} = \eta\beta^{-1} \sqrt{n_{\rm p} W \Delta\nu} \left(\frac{S_{\rm peak}}{S_{\rm sys}}\right),
\end{equation}
where $S_{\rm peak}$ is the peak detected flux density of the single pulse, $S_{\rm sys}=T_{\rm sys}/G$ is the system equivalent flux density, $T_{\rm sys} = T_{\rm rec} + T_{\rm sky}$ is the system temperature, $G$ is the system gain, $\Delta\nu$ and $n_{\rm p}$ are as defined in eq.~\ref{eq:snr_to_flux}, $W$ is the pulse width in seconds, $\eta=0.868$ accounts for an assumed Gaussian pulse shape, and $\beta\sim 1$ is a telescope degradation factor \citep[e.g.,][]{mc03}. }

\added{In our simulation, we perform this S/N calculation for each pulsar in the simulated population and find the difference between the 278 detections observed by PKMBS and the simulations, $\Delta N_{\rm detected}=N_{\rm Parkes}-N_{\rm sim}$, where $N_{\rm Parkes}$ is the number of unique single burst pulsars and RRATs detected by the PKMBS. We then vary $\sigma_{\sigma_L}$ and $\mu_{\sigma_L}$ via a grid search to optimise for $\Delta N_{\rm detected}=0$. We find the calibrated parameters are $0\leq\sigma_L\lesssim 1.3$ and $0\leq\sigma_{\sigma_L}\lesssim 0.8$ where they are degenerate in that high $\sigma_L$ corresponds to low $\sigma_{\sigma_L}$ and vice-versa.\\}

\added{Applying these calibrated parameters to \chimefrb{}, we estimate that 850--1550 single-pulse emitting objects could be detected in a two year time span, including known pulsars. We then compare simulations (570--960 pulsars/RRATs) to CHIME/FRB metadata detections of candidates and known pulsars ($\sim$750) and find that the simulated numbers are in agreement with the number of unique pulsars detected over a 90 day period. }

\added{However, this number is likely an overestimate due to our limited knowledge of the single pulse statistics of both RRATs and pulsars. In particular, the spectral index distribution may be wider rather than the narrow Gaussian given in \cite{ly95}, and our single pulse luminosity distribution are calibrated against a largely ``regular'' pulsar population from PKMBS (only 17 of the 278 were RRATs). \chimefrb{} is in a unique position to be able to characterise these single pulse emissions, so future single pulse studies with \chimefrb{} will improve our population estimates.} Given our estimate we note that we currently have a larger sample of discovered pulsars on the public webpage \footnote{\url{https://www.chime-frb.ca/galactic}.}, as well as more unconfirmed potential candidates.

\subsection{Discussion of results}
Although not primarily a pulsar discovery instrument, \chimefrb{} is a \deleted{a} useful tool in detecting pulsars and RRATs using single pulse analyses. This method samples a distinctly different section of parameter space than conventional pulsar searching methodologies. As the initial detection method does not incorporate any folding, this method can detect only sources which are bright enough to be seen at least occasionally by \chimefrb{} in single pulses. However, \chimefrb{} passively observes the \deleted{entire} Northern sky \added{above approximately declination $-15\degr{}$} continuously; this suggests that if a source emits sufficiently brightly during the period of time when it is visible to the system, we will detect that emission even if it is infrequent. Though \chimepsr{} observations require intentionally including the source in the automated scheduler, these observations can also be taken on daily or near-daily cadences. Very high cadence observations allow us to materially increase the chances that we observe an intermittent source when it is active. \added{Though single pulse searches are common, daily cadence, full-sky single pulse searches are not.}

The success of \chimefrb{} at detecting highly intermittent sources suggests that other FRB searches, e.g., ASKAP's CRAFT survey \citep{jbm+19} or MeerKAT's MeerTRAP \citep{s16} could be successfully applied to this task. \chimefrb{} has a particularly wide field of view and is fortunate to be guaranteed daily observations, but even FRB search instruments less optimized for this method could use metadata from Galactic detections to seed targeted pulsar searches.

\section{Conclusions \& Future Work \label{sec:conclusion}}
The discovery of four new RRAT-like sources and three new pulsars represents only a starting point for detecting intermittent sources with \chimefrb{} and \chimepsr{}. Simulation work conducted with PsrPopPy altered for single bursts suggests that we could observe \replaced{thousands}{hundreds} of sources using \chimefrb{} single pulse triggers. Though these simulation results are preliminary and likely an overestimate, we do expect further detections with \chimefrb{} and are currently examining other candidate sources. We will improve these simulations in the future by using the \chimefrb{} injection system. 

In this work, we focus only on reporting initial detections and solutions for these sources, but \replaced{it is clear that there is more scope for inquiry within these seven sources.}{there is further potential within these seven sources.} \added{ For example, continued observations may allow us to determine a spin period for PSR J1931+42 or may allow us to determine the nature of the changing morphology in PSR J0121+53.} As we continue to detect more sources using \chimefrb{}, we will be able to better constrain the population of RRATs and pulsars detectable with single pulses.

The possibility of long-term monitoring of these sources and further sources we will likely detect with \chimefrb{} opens doors in studying intermittent pulsars and RRATs. In more systematic future work, we may be able to better understand the continuum between persistent and intermittent emission, including intermittent pulsars and RRATs. 

These detections serve as a reminder of \chimepsr{}'s potential as a monitor for known RRATs. A systematic program follow-up program for known RRATs with \chimepsr{} may help us better understand the RRAT population as well as our new sources' position within that population. At present, it is difficult to compare our new sources to the existing population of RRATS and \added{nulling or intermittent pulsars}, but further discovery of new sources and study of existing RRATs with \chimepsr{} will allow us to add new constraints to our understanding of population.

\chimepsr{} and \chimefrb{} are key instruments in the future of FRB discovery and known pulsar studies, but together they also demonstrate that instruments which have finding FRBs as their primary science goal have the potential to generate a renaissance in the study of intermittent pulsars and nulling pulsars that have so far been difficult to discover in pulsar searches.

\acknowledgements{\added{We acknowledge that CHIME is located on the traditional, ancestral, and unceded territory of the Syilx/Okanagan people.} The CHIME/FRB Project is funded by a grant from the Canada Foundation for Innovation 2015 Innovation Fund (Project 33213), as well as by the Provinces of British Columbia and Qu\'{e}bec, and by the Dunlap Institute for Astronomy and Astrophysics at the University of Toronto. Additional support was provided by the Canadian Institute for Advanced Research (CIFAR), McGill University and the McGill Space Institute via the Trottier Family Foundation, and the University of British Columbia. The Dunlap Institute is funded by an endowment established by the David Dunlap family and the University of Toronto. Research at Perimeter Institute is supported by the Government of Canada through Industry Canada and by the Province of Ontario through the Ministry of Research \& Innovation. The National Radio Astronomy Observatory is a facility of the National Science Foundation operated under cooperative agreement by Associated Universities, Inc. We thank Erik C. Madsen for his work in planning the development of the CHIME/Pulsar backend. We are grateful to the staff of the Dominion Radio Astrophysical Observatory, which is operated by the National Research Council Canada. Pulsar research at UBC is funded by a Natural Sciences and Engineering Research Council (NSERC) Discovery Grant and by the Canadian Institute for Advanced Research (CIFAR). This research was enabled in part by support provided by WestGrid (\url{www.westgrid.ca}) and Compute Canada (\url{www.computecanada.ca}). D.C.G is supported by the John I. Watters Research Fellowship. K.C. is a UBC Four Year Fellow. V.M.K. holds the Lorne Trottier Chair in Astrophysics \& Cosmology and a Distinguished James McGill Professorship and receives support from an NSERC Discovery Grant and Herzberg Award, from an R. Howard Webster Foundation Fellowship from the CIFAR, and from the FRQNT Centre de Recherche en Astrophysique du Qu\'{e}bec. S.M.R. is a CIFAR Fellow and is supported by the National Science Foundation (NSF) Physics Frontiers Center award 1430284. M.D. receives support from a Killam fellowship, NSERC Discovery Grant, CIFAR, and from the FRQNT Centre de Recherche en Astrophysique du Qu\'{e}bec. P.S. is a Dunlap Fellow and an NSERC Postdoctoral Fellow. B.M.G. acknowledges the support of the Natural Sciences and Engineering Research Council of Canada (NSERC) through grant RGPIN-2015-05948, and of the Canada Research Chairs program. P.C. is supported by an FRQNT Doctoral Research Award. M.B. is supported by an FRQNT Doctoral Research Award.}

\software{\dspsr{} \citep{dspsr}, \presto{} \citep{presto}, \psrchive{} \citep{psrchive04, psrchive12}, \mbox{PSRPopPy} \citep{blr+14}, PulsePortraiture \citep{p19}, \tempo{} \citep{tempo}, \tempotwo{} \citep{tempo2a}}

\bibliography{references}

\end{document}